\documentclass{aa}  

\usepackage{txfonts}
\usepackage[export]{adjustbox}
\usepackage{subcaption}
\usepackage[colorlinks,linkcolor=blue,anchorcolor=blue,citecolor=blue]{hyperref}
\usepackage{xcolor}
\newcommand{\uint}{$\rm mJy\,beam^{-1}$}
\newcommand{\msd}{$\Sigma_{\rm tot}$}
\newcommand{\sfrd}{$\Sigma_{\rm SFR}$}
\newcommand{\usfr}{$\rm M_{\sun}\,yr^{-1}$}
\newcommand{\usfrd}{$\rm M_{\sun}\,yr^{-1}\,kpc^{-2}$}
\newcommand{\umsd}{$\rm M_{\sun}\,kpc^{-2}$}
\newcommand{\ussfr}{$\rm yr^{-1}$}

\begin{document} 

    \title{CHANG-ES}

   \subtitle{XXXVI. The thin and thick radio discs}
   \author{V. Heesen\inst{1}
          \and
          M. Stein\inst{2}
          \and 
          N. Pourjafari\inst{3}
          \and
          M. Br\"uggen\inst{1}
          \and
          J. Stil\inst{3}
          \and
          J.-T. Li
          \inst{4}
          \and
          T. Wiegert\inst{5}
          \and
          J. Irwin\inst{6}
          \and
          R.-J. Dettmar\inst{2}
          \and 
          T.~A.~Porter\inst{7}
          \and
          Y.~Stein\inst{2}\thanks{Now at Deutsches Zentrum f\"ur Luft- und Raumfahrt e.V. (DLR), 53227 Bonn, Germany.}
         }

   \institute{Hamburg University, Hamburger Sternwarte, Gojenbergsweg 112, 21029 Hamburg, Germany\\
              \email{volker.heesen@uni-hamburg.de}
        \and 
        Ruhr University Bochum, Faculty of Physics and Astronomy, Astronomical Institute (AIRUB), 44780 Bochum, Germany
        \and
        Department of Physics and Astronomy, University of Calgary, Canada
        \and
        Purple Mountain Observatory, Chinese Academy of Sciences, 10 Yuanhua Road, Nanjing 210023, China
        \and
        Instituto de Astrofısica de Andalucıa (IAA-CSIC), Glorieta de la Astronomıa, 18008, Granada, Spain
        \and
        Department of Physics, Engineering Physics, \& Astronomy, Queens University, Kingston, ON K7L 3N6, Canada
        \and 
        W.W. Hansen Experimental Physics Laboratory and Kavli Institute for Particle Astrophysics and Cosmology, Stanford University, Stanford, CA 94305, USA
       }

   \date{Received 6 Feb 2025 / Accepted 15 May 2025}
   
 
  \abstract
   {Edge-on spiral galaxies give us an outsiders' view of the radio halo, which envelops these galaxies. The radio halos are caused by extra-planar cosmic-ray electrons that emit synchrotron emission in magnetic fields.}
   {We aim to study the origin of radio halos around galaxies and infer the role of cosmic-rays in supporting the gaseous discs.  We would like to test the influence of star formation as the main source of cosmic rays as well as other fundamental galaxy properties such as mass and size.}
   {We present a study of radio continuum scale heights in 22 nearby edge-on galaxies from the CHANG-ES survey. We employ deep observations with the Jansky Very Large Array in the $S$-band (2--4\,GHz), imaging at $7\arcsec$ angular resolution. We measure scale heights in three strips within the effective radio continuum radius, correcting for the influence of angular resolution and inclination angle. We include only galaxies where a distinction between the two disc components can be made in at least one of the strips, providing us with robust measurements of both scale heights.}
   {We find a strong positive correlation between scale heights of the thin and thick discs  and star-forming radius as well as star-formation rate (SFR); moderately strong correlations are found for the mass surface density and the ratio of SFR-to-mass surface density; no correlation is found with SFR surface density alone. Yet the SFR surface density plays a role as well: galaxies with high SFR surface densities have a rather roundish shape, whereas galaxies with little star formation show only a relatively small vertical extent in comparison to their size.}
   {Thick gaseous discs are partially supported by cosmic-ray pressure. Our results are a useful benchmark for simulations of galaxy evolution that include cosmic rays.}


\titlerunning{The thin and thick radio disc in galaxies}
\authorrunning{V.~Heesen et al.}

   \maketitle
%

\section{Introduction}

In the present Universe, disc galaxies consist of both a thin and a thick disc. The thin disc is composed of stars, dust, and gas with a scale height of a few hundred parsecs, respectively \citep{ferriere_01a}. The thick disc comprises mostly ionised gas, in contrast to the thin disc where neutral and molecular components dominate, cosmic rays, and magnetic fields. While the thin disc is supported by the thermal pressure of the gas, the thick disc is inflated by non-thermal pressures \citep{cox_05a}. The thick stellar disc has a different kinematic and chemical composition than the thin disc and thus can be used to constrain models for galaxy formation \citep{tsukui_25a}. The thick gaseous disc is the place where we can probe for outflows and inflows, and it therefore acts as an interface between the galaxy and the circumgalactic medium which harbours a large reservoir of baryons surrounding galaxies \citep{tumlinson_17a}. In addition to the thin and thick discs, we note the existence of some more extended structures, such as the large-scale bubble discovered NGC\,4217 \citep{heesen_24a} and the extended radio blob around NGC\,5775 \citep{li_08a, heald_22a}. It is unclear whether such large scale radio features ($>$10\,kpc) exist only in special cases, however, it is worth mentioning that there may be some more extended features beyond the thin and thick discs.

Observations of galaxies in the radio continuum allow us to study the distribution of cosmic rays and magnetic fields. Both are intertwined and often times difficult to separate, yet synchrotron emission is one of the few avenues to study their influence on galaxy evolution. This is an ongoing research topic as comic rays follow magnetic field lines, and they are theorised to have a crucial effect on the efficiency of stellar feedback \citep{ruszkowski_23a}. This may be surprising as cosmic rays receive only 10\,\% of the kinetic energy in the shock waves of supernova remnants, with the remainder thermalised to heat the gas, but there appears to be an increasing understanding of the theoretical underpinnings of winds that are driven by cosmic rays \citep{thompson_24a}. Briefly, cosmic rays are transported out of their sites of injection in star-forming regions, and thus establishing a vertical pressure gradient. This gradient accelerates gas away from galaxies without heating it \citep{salem_14a}.

One of the aims of the Continuum HAloes in Nearby Galaxies - an EVLA Survey (CHANG-ES) project was to investigate the workings of these processes by observational means \citep{irwin_12a,irwin_24a}. CHANG-ES is a radio continuum survey of 35 nearby galaxies including $L$- (1--2\,GHz) and $C$-band (4--6\,GHz) data with full polarisation. CHANG-ES has so far delivered many insights into the structure and dynamics of the extra-planar interstellar medium. In combination with observations from the LOw Frequency ARray \citep[LOFAR;][]{vanHaarlem_13a}, detailed observations of spectral ageing made it possible to study the transport of cosmic-ray electrons from the disc into the halo \citep{schmidt_19a,stein_19a,stein_19b,miskolczi_19a,heald_22a,stein_23a}. These results were underpinned by a scale height analysis of the entire sample that showed that the escape of cosmic-ray electrons is possible, so that the galaxies are non-calorimetric \citep{krause_18a}. New insights into the magnetic field structure have also been gained. In NGC\,4631, using Faraday rotation, magnetic ropes were revealed that have sizes of kiloparsecs with alternating magnetic field directions along the line of sight \citep{mora_19a}. In NGC\,4666, a quadrupolar magnetic field structure was detected implying that there is a reversal of magnetic field direction across the plane of the disc \citep{stein_19a}. Such a symmetry would be expected from the existence of a large-scale galactic dynamo operating at scales of tens of kiloparsecs, equivalent to the size of the galaxy. \citet{henriksen_21} developed a theoretical basis for such large-scale field patterns. Finally, similar results were confirmed by a Faraday rotation study of the entire sample that shows the prevalence of X-shaped magnetic fields that are likely connected to galactic winds \citep{krause_20a} with a corresponding study of the X-shaped magnetic field morphology presented in \citet{stein_25a}.

With new $S$-band data we are enriching the existing survey. With new deep polarimetric $S$-band (2--4\,GHz) C-configuration observations our aim is to cover a broad range of galaxy properties and strengthen the overall analysis. $S$-band polarised emission is the best probe for magnetic fields in the galaxies' faint halos, and $S$-band is located in the crucial transition between Faraday-thin and -thick frequency regimes.  While the CHANG-ES $L$-band data suffer from strong depolarisation and the $C$-band data have limited resolution in Faraday space, our newly observed $S$-band data are essential, and will, in combination with the existing $C$- and $L$-band data, provide new insights on magnetic halos of galaxies. Applying rotation measure synthesis to the combined data from $L$, $S$ and $C$ bands will achieve an unprecedented resolution in Faraday space, and letting us map the 3-D structure of the ordered magnetic field. The results will significantly improve our understanding of extended halo structures of a variety of magnetic field patterns in galaxies. First results were presented by \citet{heesen_24a}, \citet{irwin_24b}, and \citet{xu_25a} showing the versatility of the new $S$-band data.

In this new study we are revisiting some of the work presented by \citet{krause_18a}. They analysed the radio continuum scale height of the thick radio disc in both $L$ and $C$ bands. While they modelled the vertical intensity distribution with two components, they did not include the analysis of the thin disc. We are extending their work in two important ways. First, we include the analysis of the thin radio disc; second, we calculate deprojected mean intensities in both the thin and thick disc \citep{stein_19b,stein_20a}. We use $S$-band data because of its superior angular resolution and sensitivity. A potential study of the integrated spectral behaviour of both disc components is deferred to later work. In this work, we will only use the total power radio continuum maps. We considered for analysis 22 galaxies from the CHANG-ES sample, properties of which can be found in Table\,\ref{tab:sample}. A full presentation of the data including polarisation will be presented in a forthcoming paper (Wiegert et al., in prep.).

This paper is organised as follows: Section\,\ref{s:data_and_methodology} describes our sample selection, data reduction, and methodology. In Sect.\,\ref{s:results} we present our resulting scale heights and flux densities, which we relate to fundamental galaxy parameters. We discuss our results in Sect.\,\ref{s:discussion} before we conclude in Sect.\,\ref{s:conclusions}. Appendix\,\ref{as:sample} contains more information about our galaxy sample and Appendix\,\ref{as:results} presents our results in more detail.

\begin{figure*}
    \centering
    \includegraphics[width=\linewidth]{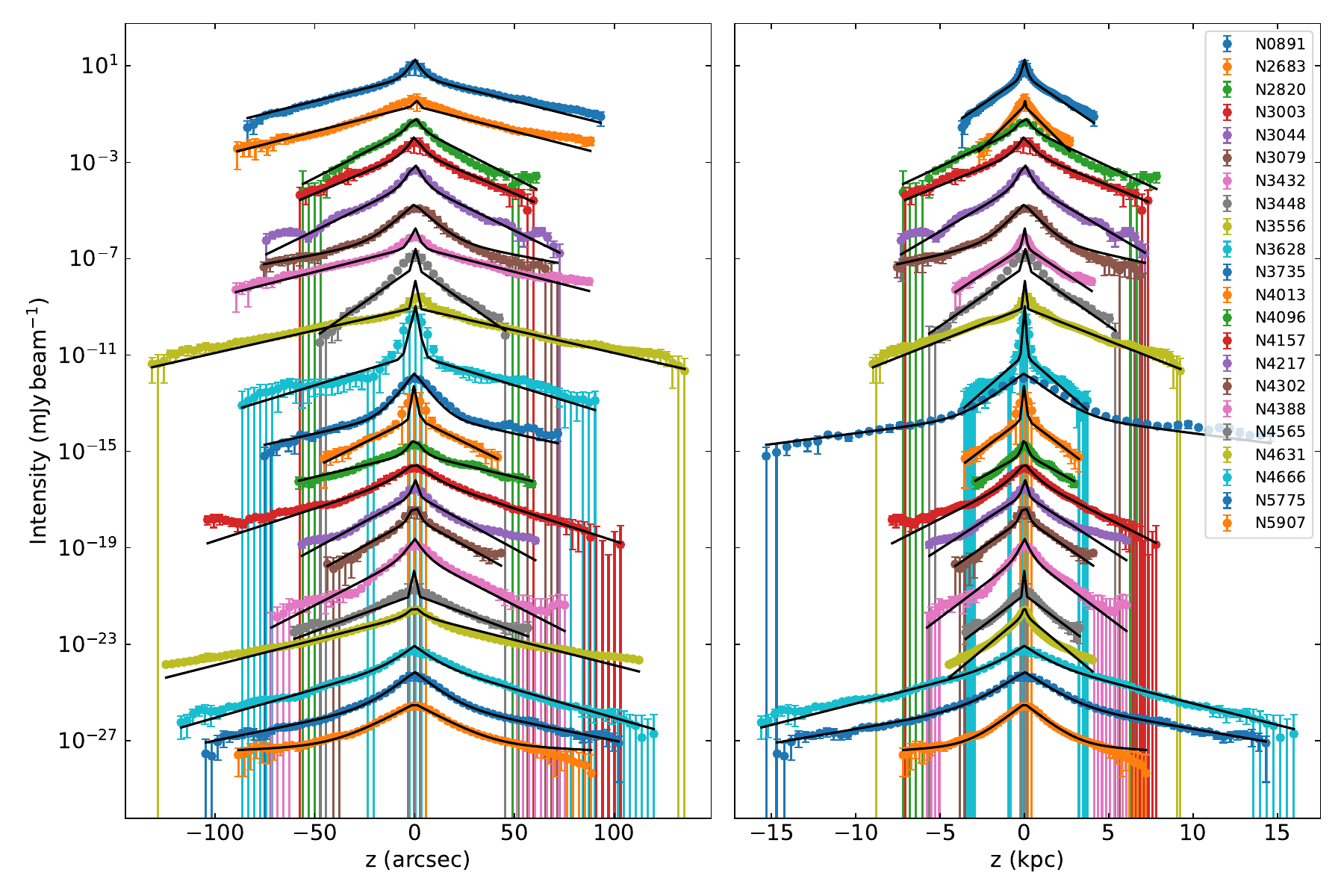}
    \caption{Vertical radio continuum intensity profiles at 3\,GHz in our sample galaxies. We show intensities in the central strip of each galaxy, with exception of NGC\,3079, 4388, and 4666 where those in the eastern strip are shown. The left panel shows them as observed on the sky, whereas the right panel shows them projected to the assumed distance. Solid lines show two-component exponential model profiles deconvolved from the effective beam. The intensities were arbitrarily scaled in order to separate the profiles for improved display.}
    \label{fig:int}
\end{figure*}

\section{Data and methodology}
\label{s:data_and_methodology}

\subsection{Observations}
\label{s:observations}

Observations were taken with the JVLA in the $S$-band in the frequency range of 2--4\,GHz using full polarisation as part of observing programs 20A-018, 21A-033, and 22B-016 (PI: Y.~Stein). We used the C-configuration which provides us with a nominal resolution of around $7\arcsec$ using Brigg's robust weighting. The field of view as given by the diameter of the primary beam is $15\arcmin$. The largest angular scale that we can image is $8\farcm 2$. Observations were taken in the standard fashion with 2048 channels of 100\,kHz bandwidth each. We began with a scan of the primary calibrator used for flux scale and bandpass calibration. Our target observations are bracketed in scans of the secondary calibrator to determine the phase calibration.

\subsection{Data reduction}
\label{ss:data_reduction}

Our data are reduced with the Common Astronomy Software Applications \citep[{\sc casa};][]{casa_22a} version {\sc pipeline} 6.4.12. For Stokes $I$ we used the NRAO pipeline without any further calibration or data flagging. The data were imaged with {\sc wsclean} version $2.9$ using a multi-scale {\sc clean} algorithm \citep{offringa_14a}. The $S$-band data have a field of view of $15\arcmin$, so that we have made maps with a size of two times the primary beam extent. For a few targets, we created larger maps in order to deconvolve confusing sources beyond a distance of one primary beam diameter. We then performed two rounds of self-calibration in phase only before creating final maps. Only for a few targets with strong residuals we also performed one round of self-calibration in phase and amplitude. Also, for several sources we used mosaicked observations since they have a large apparent size. These are NGC\,891, 3628, 4565, and 4631. In this case (with the exception of NGC\,891, see below), we treated each of the two pointings separately and then combined them in a linear fashion using {\sc linmos} within {\sc casa}. 

We imaged and deconvolved maps using Brigg's robust weighting scheme with {\tt robust} parameter set to $0.5$. Final maps were restored with Gaussian {\sc clean} beams using $7\arcsec$ full width at half maximum (FWHM). We also applied the primary beam correction with {\sc wsclean}. Typical rms map noise values are between $3.5$ and $5\,\muup\rm Jy\,beam^{-1}$. 

For NGC\,891 we used joint deconvolution with {\sc tclean} as part of {\sc casa}. Two observations of NGC\,891 were conducted in the $S$-band, each comprising two pointings. A single phase-only self-calibration loop was performed on the first observation, while two amplitude and phase self-calibration loops were applied to each observation.  The data were manually flagged to remove any radio frequency interference (RFI). These data were imaged using the multi-scale {\sc tclean} algorithm, with the {\tt gridder} parameter set to {\tt mosaic} to account for the multiple pointings and Brigg's robust weighting of zero (${\tt robust} = 0$; Pourjafari et al. in prep.). The resulting map has an angular resolution of $5\arcsec$ which we smoothed to a matching resolution.

\subsection{Intensity profiles}
\label{ss:intensity_profiles}

We created vertical intensity profiles by averaging in boxes aligned in strips perpendicular to the major axis \citep[see e.g.][for details of this procedure]{krause_18a}. We used three strips with the middle one centred on the galaxy and the other ones symmetrically at larger galactocentric distances. For the width of the combined strips, we used the effective ($e$-folding) diameter in the radio continuum. We measured this by placing a $10\arcsec$-wide strip centred on the major axis and fitting a Gaussian profile to the radio continuum emission. Hence, the width of each vertical strip is $2/3 r_{\rm e}$, where $r_{\rm e}$ is the effective radius in the radio continuum. Intensities are then averaged in boxes with the width equivalent to strip width and with a height of approximately half the angular resolution. The height of each box was chosen at $3\arcsec$. The intensity profiles were created using the {\sc nod3} software package, where we made use of the {\sc boxmodels} function \citep{mueller_17a}. 

In Fig.\,\ref{fig:int} we present the vertical intensity profiles in our galaxies in an example (mostly the middle) strip. In the linear--log $xy$ diagram an exponential function is a straight line. However, one can clearly see that more than one component is required because the slope of the profiles changes perceptively near the galactic mid-plane. In some cases, such as in NGC\,891, a clear break is visible, in other cases the change is more gradual such as in NGC\,5775. This means we can observe both the thin and thick discs and are able to separate them. 

We fitted exponential functions to the intensity profiles. Because of the limited angular resolution, the influence of the effective beam has to be taken into account. The effective beam is the superposition of the {\sc clean} beam (i.e.\ the intrinsic angular resolution) and the contribution from the inclined disc. Because our galaxies are not perfectly edge-on, the projected contribution from the radial profile along the minor axis needs to be considered. The resulting theoretical profiles are thus the convolution of the exponential distribution with a Gaussian function describing the effective beam, which  can be described by an analytic formula \citep{dumke_95a}. In Fig.\,\ref{fig:int} we show for each galaxy a model intensity profile for the two-component exponential distribution that are deconvolved from the effective beam. For our final results we took the weighted mean of the amplitudes and scale heights of the thin and the thick discs in those strips where the reduced chi-squared value fulfilled $\chi_\nu^2< 3$. In some cases it was necessary fit both haloes individually as the intensity profiles are asymmetric; in other cases only one halo (either northern or southern) was fitted. The resulting amplitudes and scale heights of the thin and thick disc are presented in Table\,\ref{tab:results}.

To fit the intensity profiles with the two-component exponential profiles, {\sc nod3} uses the {\sc SciPy} \citep{scipy_20a} {\sc curve\_fit} implementation of the Levenberg--Marquardt algorithm and performs the uncertainty estimation for the fitting parameters based on the covariance matrix. Such an approach can lead to an underestimation of the parameter uncertainties, especially in the case of correlated parameters. Therefore, we refitted individual intensity profiles using a Markov chain Monte Carlo approach and found overall agreeing results. Based on these tests, we decided to use the fitting routine as implemented by {\sc nod3}.

\subsection{Integrated intensity profiles}

We have integrated the intensities along the vertical direction in order to derive integrated flux densities \citep{stein_19a}. These can be deprojected using the effective radius in radio continuum emission, so that we obtain a mean deprojected intensity as it would appear in a face-on galaxy. Hence
\begin{equation}
    I_i = w_i\,z_i/r_{\rm e},
\end{equation}
where $I_i$ is the deprojected mean intensity, $w_i$ is the amplitude in units of $\rm mJy\,beam^{-1}$,  $z_i$ is the scale height in units of $\rm arcsec$, and $r_{\rm e}$ is the effective radius in units of $\rm arcsec$. Here, $i\in \lbrace 1,2\rbrace$ is for the thin and thick disc, respectively.  The mean intensity of the combined thin and thick radio disc is then $I=I_1+I_2$.

\subsection{Sample}
\label{s:sample}

The CHANG-ES sample consists of 35 galaxies. These were all included in the $S$-band survey with the exception of NGC\,4244. In our analysis we included only galaxies that allowed us to separate the thin and thick disc contributions. This was decided on the basis of the reduced chi-squared value $\chi_\nu^2$ of the fitted intensity profiles. Only strips with $\chi_\nu^2\leq 2$ were considered acceptable for the analysis. In those cases where a single component resulted in acceptable fits, we excluded the galaxy as the thin and thick discs could not be separated (NGC\,3877, NGC\,4192, UGC\,10288). In some cases the projected disc was so prominent that no suitable profiles for fitting could be obtained (NGC\,2613, 5297, 5792). Also, galaxies that showed only nuclear emission were omitted (NGC\,2992, 4594, 4845, and 5084) as well as strongly interacting galaxies (NGC\,660, 4438). This leaves us with a sample of 22 galaxies, properties of which can be found in Table\,\ref{tab:sample}. 

\begin{figure}
    \centering
    \includegraphics[width=\linewidth]{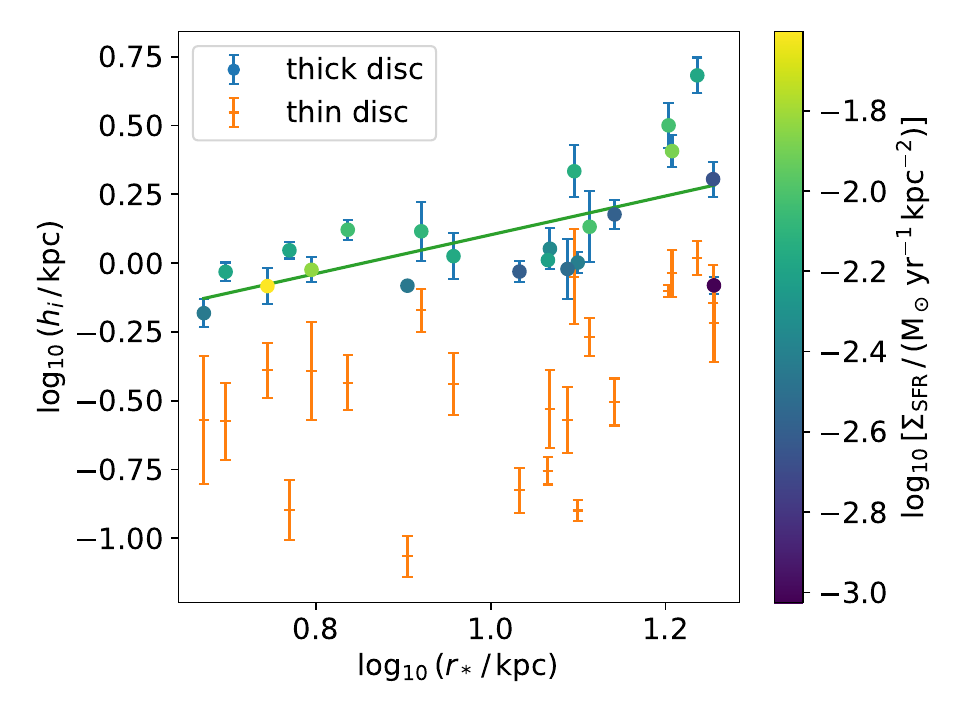}
    \caption{Scale height as function of star-forming radius. For the thick disc the best-fitting relation is shown. Data points for the thick disc are colour-coded with the SFR surface density.}
    \label{fig:sh_radius}
\end{figure}
Since our main results are based on the decomposition of the thin and thick disc components, we may need to discuss a possible bias caused by the inclination correction. This could be particularly important for the thin disc component. This is because the selection of our sample is based on if we can efficiently decompose these two components. We found no correlation between inclination angle and scale height ($\rho_{\rm s}\leq 0.1$). We chose inclination angles of \citet{krause_18a}, or otherwise from \citet{irwin_12a}. The former ones were obtained from scale height fitting similar to ours, the later ones are from optical morphology. For an assumed inclination angle uncertainty of $5\degr$ we estimate that the uncertainty of the thin disc is up to 50\,\%, whereas that of the thick disc is less than 20\,\%. 

\begin{table}
\caption{Best-fitting relations using the equation $\log_{10}(y) = a\log_{10}(x) + b$ with Spearman's rank correlation coefficient $\rho_{\rm s}$.}
\label{tab:fit}
\centering
\begin{tabular}{l c c c c}
\hline\hline 
$y$    & $a$ & $b$ & $x$ & $\rho_{\rm s}$ \\
\hline
$h_2$  &     $0.70\pm 0.20$ & $-0.60\pm 0.21$ & $r_\star$ & $0.63$ \\ 
$h_2$  &     $0.43\pm 0.07$ & $-0.002\pm 0.035$ & $\rm SFR$ & $0.70$ \\ 
$h_2$  &     $0.15\pm 0.15$ & $0.44\pm 0.34$ & \sfrd & $0.27$ \\
$h_2$  &     $-0.33\pm 0.21$ & $2.82\pm 1.75$ & \msd & $-0.43$ \\
$h_2$  &     $0.33\pm 0.14$ & $3.56\pm 1.47$ & \sfrd/\msd & $0.45$ \\
$I_1$  &     $1.20\pm 0.30$ & $1.61\pm 0.68$ & \sfrd & $0.68$ \\
$I_2$  &     $0.89\pm 0.36$ & $0.73\pm 0.81$ & \sfrd & $0.38$ \\
$I$    &     $1.10\pm 0.25$ & $1.68\pm 0.58$ & \sfrd & $0.63$ \\
\hline               
\end{tabular}
\end{table}

\begin{figure*}
    \centering
    \includegraphics[width=\linewidth]{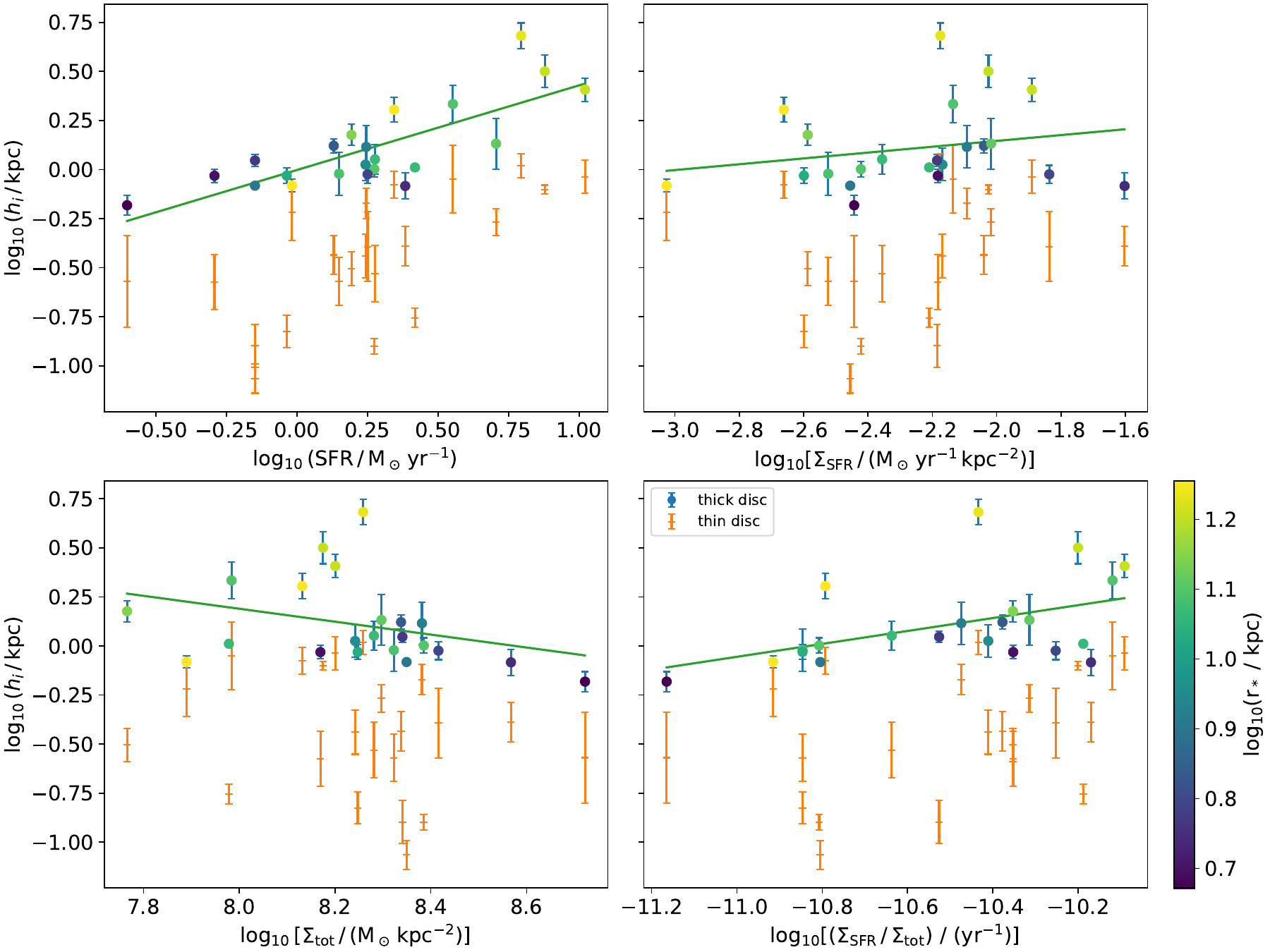}
    \caption{Scale height in the thin and thick radio disc as function of various galaxy parameters. In each panel we show the data points for the thick disc as filled circles colour-coded with star-forming radius. Also, the best-fitting line for the thick disc is shown. The individual panels show the scale height as a function of SFR (top left), SFR surface density (top right), mass surface density (bottom left), and ratio of SFR-to-mass surface density (bottom right).}
    \label{fig:sh}
\end{figure*}

\section{Results}
\label{s:results}

\begin{figure}
    \centering
    \includegraphics[width=\linewidth]{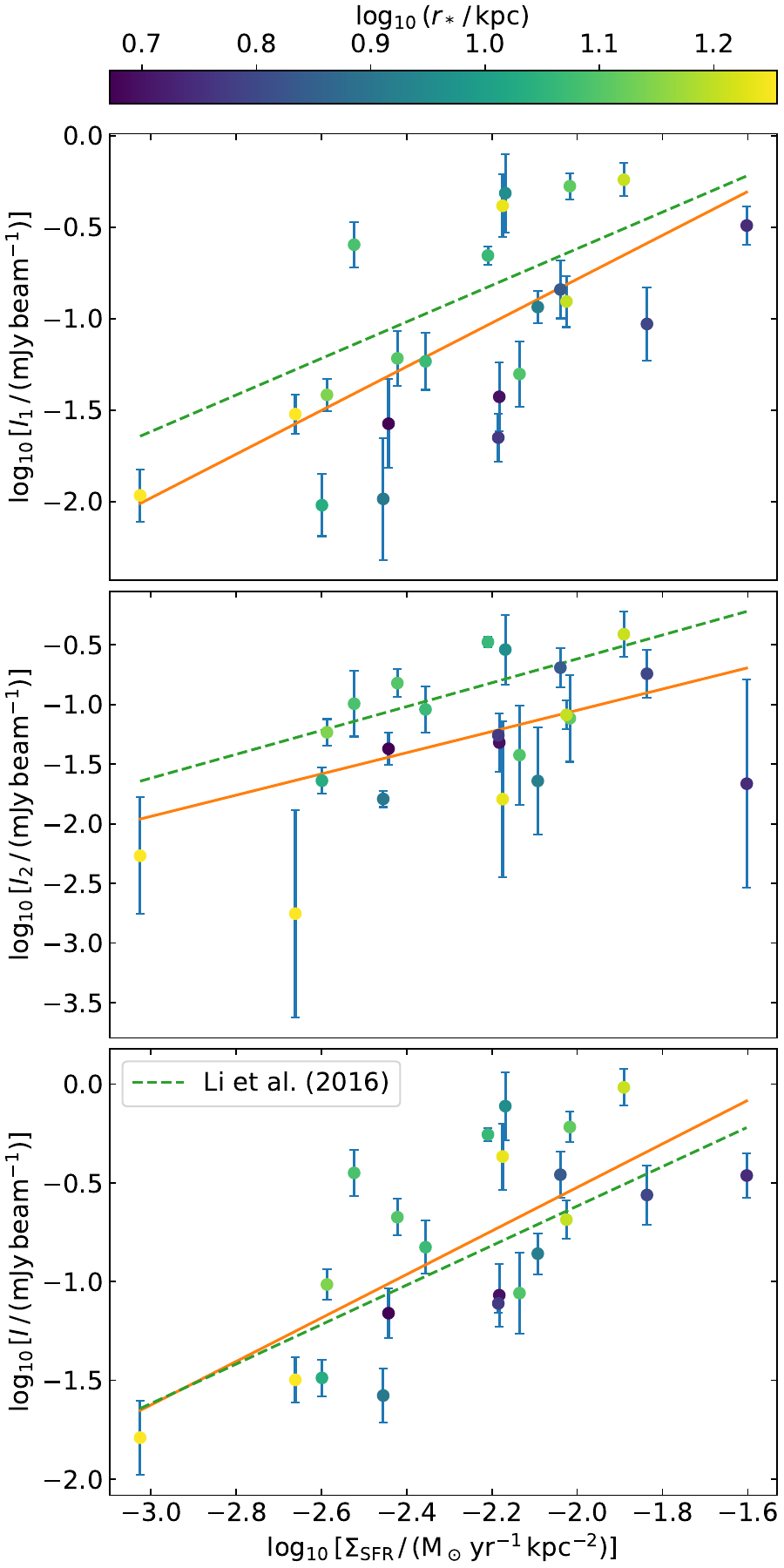}
    \caption{Deprojected mean intensities as function of SFR surface density. Best-fitting relations are shown as solid lines, and  the relation for integrated luminosities is shown as dashed line \citep{li_16a}. We show the relation in the thin disc (top panel), thick disc (middle panel), and combined thin and thick disc (bottom panel).}
    \label{fig:l_sfrd}
\end{figure}

\subsection{Scale heights}
\label{ss:scale_heights}

First, we investigate the scale heights in both the thin and the thick discs. We find arithmetic mean scale heights of $0.45\pm 0.06$ and $1.47\pm 0.20$\,kpc of the thin and thick disc (see Table\,\ref{tab:results}, for values), respectively. In Fig.\,\ref{fig:sh_radius} we show the dependence of scale height as a function of star-forming radius. We find significant correlations ($p$-value of $p<0.05$ from Spearman's test) between both thin and thick disc scale height and radius; both have moderate correlation strengths (Spearman's correlation coefficient of $0.3\leq  \rho_{\rm s} \leq 0.6$). Information about the best-fitting relation as shown in Fig.\,\ref{fig:sh_radius} and of those in the following figures is presented in Table\,\ref{tab:fit}. A related positive correlation of the thick disc scale height with radio disc diameter was found by both \citet{krause_18a} and \citet{galante_24a}. Figure\,\ref{fig:sh_radius} also shows that points above the best-fitting relation have predominantly high star-formation rate (SFR) surface densities, while for the ones below the opposite is true. Hence, the size of the star-forming disc alone seems to be insufficient to determine the size of the radio halo, but a sufficient level of star formation may also be required. 

Hence, we expect a combination of SFR surface density and radius, such as the SFR, to be important in determining the scale height. This is indeed the case we show in Fig.\,\ref{fig:sh}. Both for the thin and the thick disc, we found significant and strong positive correlations ($p<0.05$, $\rho_{\rm s}> 0.6)$. Moreover, we investigated the dependence on SFR surface density alone. In this case we find no significant correlation for the thick disc ($p>0.1$, $\rho_{\rm s}<0.3$) and a moderate positive correlation for the thin disc. One reason may be that the influence of the star-forming radius is too large, as galaxies below the best-fitting relation have smaller radii while the opposite holds for galaxies above the relation. A positive correlation with SFR surface density may be expected if the increased stellar feedback leads to outflows of cosmic rays and magnetic fields from the star-forming disc. If that is the case, then we also may expect to see a dependence on the mass surface density that governs gravitational attraction. In this case we find indeed evidence for a correlation ($p<0.05$), with a moderate negative correlation strength ($-0.6\leq \rho_{\rm s}< -0.3$). Assuming for a moment this correlation is real, and because the SFR and the mass surface densities are normalised with the same geometric area, we may consider simply the SFR-to-mass ratio instead. This quantity is related to specific star formation rate. However, because we using total instead of stellar mass, the concept is not quite applicable and we simply refer to the ratio instead. We find a significant moderate positive correlation between disc scale height and the ratio of SFR-to-mass surface density in both thin and thick discs. 

Hence, we conclude from this section that the SFR and the star-forming radius appear to be the prime parameters in determining the scale height of the radio discs. The negative correlation with the mass surface density and the positive correlation with the ratio of SFR-to-mass surface density are important, too, as secondary parameter. We also re-analysed the data in the $L$-band by \citet{krause_18a} who did not report a significant relation between thick disc scale height and SFR. We found that using the revised SFR values of \citet{vargas_19a} showed that only NGC\,3003 is an outlier in the $L$-band data, whereas NGC\,3735 \citep[the galaxy with the second-lowest SFR surface density in Fig.\,10 of][]{krause_18a}, lies on the best-fitting relation \citep[the revised SFR value of NGC\,3735 is a factor three higher, see][]{vargas_19a}. Therefore, we found that the $L$-band data shows a significant positive correlation with SFR just as we found it. However, due to the limited sample size it was not quite so obvious visually when studying the relation. Therefore, it is plausible that \citet{krause_18a} did not report it.

\subsection{Deprojected mean intensities}
\label{ss:deprojected_intensities}

In this section, we present the relation between deprojected mean intensities with the SFR surface density. In Fig.\,\ref{fig:l_sfrd} we show the intensities of the thin and thick radio discs, as well as their combination, as function of the SFR surface density. All relations are significant ($p$-value $<$$0.05$). However, whereas for the mean deprojected intensity of the thin disc as well as for the combined thin and thick disc we observe a strong correlation ($\rho_{\rm s}> 0.7$), the correlation for the thick radio disc is only moderate ($\rho_{\rm s}\approx 0.4$). The relation in the thin disc is mildly super-linear with a slope of $1.20\pm 0.30$, whereas the combined relation has the expected nearly linear slope of $1.10\pm 0.25$. This slope is in good agreement with \citet{li_16a} who found a slope of $1.09\pm 0.05$ (as interpolated from their two frequencies) in the CHANG-ES sample for global measurements. 

\begin{figure}
    \centering
    \includegraphics[width=\linewidth]{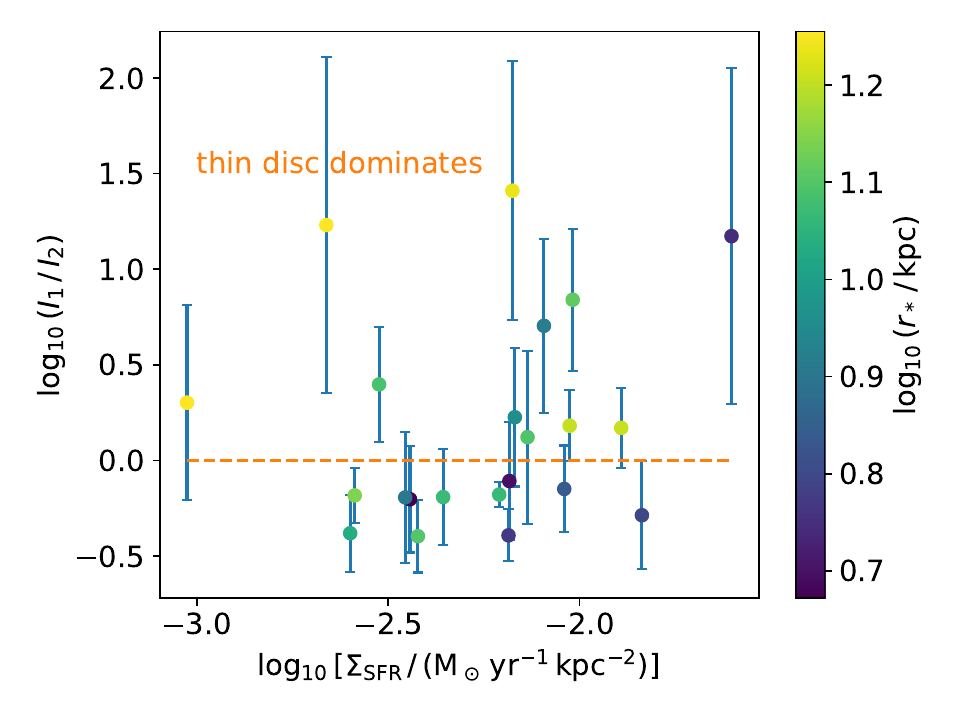}
    \caption{Ratio of thin-to-thick disc deprojected intensities. Data points are shown as function of SFR surface density and coloured according to their star-forming radius. The dashed line shows the boundary between thin disc-dominated (above line) and thick disc-dominated (below line) intensities.}
    \label{fig:disc_sfrd}
\end{figure}

In Fig.\,\ref{fig:disc_sfrd} we show the ratio of the intensities of the thin to that of the thick disc as a function of the SFR surface density. We find a no significant correlation ($p > 0.1$). The mean ratio of thin-to-thick radio intensity is $\log_{10}(I_1/I_2)=0.01\pm 0.54$. Hence, the thin radio disc contributes on average only about the same to the deprojected mean intensity as the thick radio disc. This indicates that studies of face-on galaxies may suffer from a confusion of thin and thick disc, for instance when local correlations at kiloparsec scales are studied. In earlier works it was shown that this can be partially corrected when using the radio spectral index as a secondary parameter to correct for the effects of cosmic-ray transport \citep{heesen_24a}.

\subsection{Normalised scale heights}

The geometrical size of the galaxy is important as the cosmic-ray density will decrease, both for diffusion and advection as function of source distance. The scale size for normalisation will depend on the size of the injection region. For this we may use the star formation distribution. Thus we use the effective radius as measured from the radio continuum image, to define the normalised scale height as:

\begin{equation}
    H_i = \frac{h_i}{r_e},
\end{equation}
where $r_e$ is the effective radius of the radio continuum emission.

\begin{figure}
    \centering
    \includegraphics[width=\linewidth]{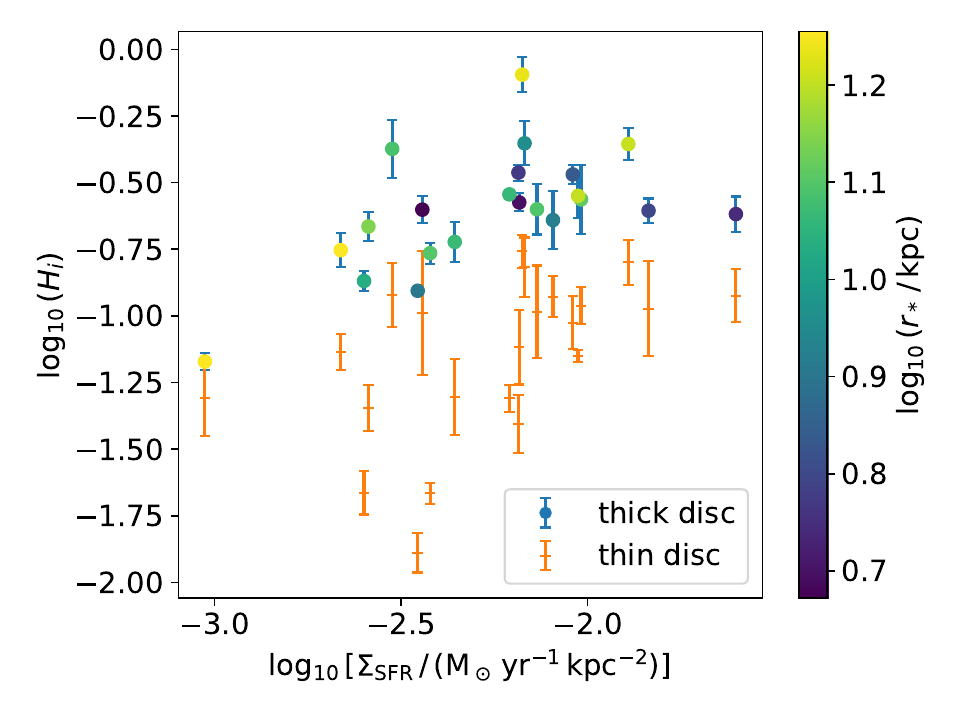}
    \caption{Normalised scale heights as function of SFR surface density. Data points of the thick radio disc are coloured according to the star-forming radius. }
    \label{fig:ratio_sfrd}
\end{figure}

Figure\,\ref{fig:ratio_sfrd} shows the normalised scale height as function of the SFR surface density. We find significant correlations for both the thin and the thick radio disc. At high SFR surface densities the scale height of the thin disc approaches a value of $0.5r_e$. This means that the shape of the galaxy becomes rounder. At low SFR surface densities in contrast, the ratio of the scale height to effective radius becomes as low as $0.1$, hence the galaxy appears to be very thin in projection. We note that for energy equipartition between the cosmic rays and the magnetic field, the cosmic-ray scale height is approximately twice as high as the synchrotron scale height. Therefore, at high SFR surface densities, the cosmic ray scale height approaches the effective radius of injection. This may give us some clues about the mode of cosmic-ray transport such as diffusion and advection.

\section{Discussion}
\label{s:discussion}

\subsection{Hydrostatic equilibrium}
\label{ss:hydrostatic_equilibium}
\citet{cox_05a} showed that cosmic rays are needed to explain the necessary pressure for a hydrostatic equilibrium in the Milky Way. Hence, we may assume:

\begin{equation}
    \frac{{\rm d}P}{{\rm d}z}= -\rho g,
\end{equation}
where $g$ is the gravitational acceleration, here assumed to be constant. We need an equation of state, such as $B\propto \rho^{0.5}$ \citep{heesen_23a} to find for the magnetic pressure and hence the cosmic-ray pressure $P_{\rm CR}=\rm const.\,\rho$. The solution is an exponential decrease $P_{\rm CR}(z) = P_{\rm CR,0} \exp{(-z/h_{\rm CR}})$ with a cosmic-ray pressure scale height of:
\begin{equation}
    h_{\rm CR}= \frac{P_{\rm CR}}{\rho g}.
    \label{eq:cosmic_ray_scale_height}
\end{equation}
We may express the gravitational acceleration via the mass surface density as $g=\piup/(2G) \Sigma_{\rm tot}$ with $G$ the gravitational constant. Hence, the scale height is proportional to the ratio of cosmic-ray pressure to density, and anti-proportional to the mass surface density. As cosmic rays are injected by star formation, a higher cosmic-ray pressure appears to be natural. Now the rate of star formation depends also on the gas density, but the slope is super-linear if the total gas density is considered \citep{kennicutt_12a}. Hence, the ratio of cosmic-ray pressure to gas density would increase with SFR surface density. Combining these two relations, the hydrostatic equilibrium predicts that the cosmic-ray scale height should be proportional to the ratio of SFR to mass surface density (Eq.\,\ref{eq:cosmic_ray_scale_height}). This is indeed what we observe, and the relation is  significant one for the scale height (Sect.\,\ref{ss:scale_heights}).

However, it is unclear if the thin disc actually is supported by the thermal pressure of the gas, while the thick disc is inflated by non-thermal pressures. There is yet little observational constraint about this. The pressure balance between different interstellar and circumgalactic medium phases was discussed in a few recent papers \citep{li_24a,li_24b}, following some simple early discussions in \cite{wang_01a} and \citet{irwin_12b}. The key conclusion is that the magnetic pressure and thermal pressure from hot gas reach a balance at a few kiloparsecs above the disc. Hence, cosmic ray pressure is expected to inflate the thick disc only.

\subsection{Radio continuum--star formation relation}

Our results for the different relations between radio continuum emission and star formation in thin and thick disc are an important test for the applicability of these relations. The global relation has a slope of close to one, when studied at gigaghertz frequencies \citep{li_16a}. At lower frequencies, we find a super-linear relation which can be explained with escape of cosmic-ray electrons in smaller galaxies \citep{smith_21a,heesen_22a}. When studied at kiloparsec scales, the relation is sub-linear, which can be explained by cosmic-ray transport. Basically, there is an excess of cosmic-ray electrons in areas of less star formation due to diffusion \citep{heesen_23a}. This relation is close to linear at gigahertz frequencies, which is usually ascribed to the fact that electron calorimetry holds in galaxies.

Now, we showed that the relation is sub-linear for the thick disc component, but super-linear for the thin disc component. We can interpret this in such a way that the young cosmic-ray electrons in the disc have not yet diffused away and therefore the relation is close to linear as also observed in face-on galaxies \citep{heesen_19a,heesen_23a}. For older cosmic-ray electrons such as in the thick disc and in the inter-arm regions, the relation is sub-linear, again as we observe. Because of this trend, there is the tendency that galaxies with higher star-formation rate surface densities have a less prominent halo in comparison to the disc. What would be interesting for future studies is to take the spectral index distribution into account. Basically, we would expect the local relation to be different from linear, as it would be a large coincidence as this depends on the magnetic field strength as well \citep{werhahn_21a}. Thus, measurements of the radio spectral index could help to identify regions where spectral ageing and low-frequency absorption \citep{gajovic_24a} plays a smaller role, so that we can measure the intrinsic relation unaffected by these effects.

As support for such future considerations, we note also that non-thermal emissions observed by the $\it Fermi$-LAT \citep{2012ApJ...755..164A} also exhibits scaling with similar measures for SFR. Modelling of Milky Way-like galaxies viewed at high inclination angles (Porter et al., in prep.) indicates that quasi-linear correlation with the infrared and radio/$\gamma$-rays holds approximately for regions where there are significant fluxes of emissions. However, it departs significantly for sight lines probing away from the geometrical thin extent of the disc. This is due to the production of the non-thermal emissions throughout the extended halo region from the presence of both cosmic rays and target distributions (magnetic field, interstellar radiation field), whereas the infrared emissions are only coming from the narrow (when viewed close to edge-on) star-forming region. Unfortunately, for our sample of galaxies, their $\gamma$-ray emissions are too weak to be detected, and none appear as sources in the latest {\it Fermi}-LAT catalogue \citep[4FGL-DR4;][]{2023arXiv230712546B}.

\subsection{Cosmic-ray diffusion}
\label{ss:cosmic_ray_diffusion}

Our findings of a positive correlation between scale height and galaxy size (Sect.\,\ref{ss:scale_heights}) can be explained as follows: cosmic rays are injected in star-forming regions and are subsequently transported away, either by diffusion or advection, with the latter similar to streaming. In the case of diffusion, the steady-state solution is \citep{quataert_22a}:

\begin{equation}
    P_{\rm CR} = P_{\rm CR,0} + \frac{\dot E_{\rm CR}}{12\pi \kappa} \left (\frac{1}{r}-\frac{1}{r_0} \right ) .
    \label{eq:diffusion}
\end{equation}
Here, $r_0$ is the base radius, $\kappa$ is the isotropic diffusion coefficient, $\dot E_{\rm CR}$ is the cosmic-ray energy injection per time, and $r$ is the radius. The base radius we have understood to be the size of the star-forming region that drives the vertical outflow. Therefore, we can associate the base radius with the effective radius of the radio continuum emission $r_{\rm e}$. Also, $P_{\rm CR,0}$ is the base cosmic-ray pressure at radius $r_{\rm e}$. With a typical cosmic-ray injection of $\dot E_c=3\times 10^{40}\,\rm (SFR / M_\sun\,yr^{-1})\,erg\,s^{-1}$ \citep{socrates_08a}, and a diffusion coefficient of $10^{28}\,\rm cm^2\,s^{-1}$, we obtain:
\begin{equation}
    P_{\rm CR} \approx 1 + r_{\rm e}^2\left ( \frac{1}{r} - \frac{1}{r_e} \right ) ,
\end{equation}
where the pressure is in units of $10^{-12}\,\rm erg\,cm^{-3}$, which we also have assumed to be the base pressure. We have assumed a SFR surface density of $10^{-2}$\,\usfrd and the radii are expressed in units of kpc. This can be approximated with an exponential function, where the cosmic-ray pressure scale height $\approx$$r_{\rm e}/4$.  This then would explain the observed positive correlation between scale height of the thick disc and star-forming radius.

\section{Conclusions}
\label{s:conclusions}

We present a study of thin and thick radio discs in a sample of star-forming, late-type edge-on galaxies. The thin disc is usually thought to be associated with the gaseous, star-forming disc with a scale height of a few hundred parsec that contains stars, dust, and neutral atomic and molecular gas. The thick disc is associated with older stars, warm and hot ionised gas, and, cosmic rays and magnetic fields. The thick disc might be the place where cosmic ray-driven winds are launched \citep{girichidis_18a}, and where the mass exchange with the circumgalactic medium happens that is so crucial for understanding galaxy evolution \citep{tumlinson_17a}. 

We use data from the CHANG-ES survey \citep{irwin_12a, irwin_24a}, exploiting the new $S$-band (2--4\,GHz) data with its superior sensitivity and angular resolution. We use a sample of 22 galaxies that fulfilled our selection criteria, in particular a clear detection of, both, thin and thick disc. We also omitted galaxies that are AGN-dominated \citep{irwin_19a} and those that are strongly interacting. We measured vertical intensity profiles in three strips centred on the minor axis, accounting for the limited angular resolution and projection effects. These profiles were then fitted with two-component exponential functions to measure the scale height and deprojected mean intensities of both the thin and thick discs.

We find mean arithmetic scale heights of the thin and thick disc of $0.45\pm 0.06$ and $1.47\pm 0.20$\,kpc, the latter value in agreement with \citep{krause_18a}. The thin disc scale height is in good agreement with the gaseous scale height of the warm neutral gas, whereas the thick disc corresponds to the warm ionised gas with a scale height in excess of 1\,kpc \citep{lu_23a}. We found that the scale heights show a strong positive correlation with the star-forming radius, as was previously reported by \citet{krause_18a}. Similarly, they show a moderate negative correlation with the mass surface density. We explored also the dependence on SFR surface density, but found no significant correlation for the thick disc. Again, both finding were already reported by \citet{krause_18a}. As the size of the galaxy is playing an important role, we then investigated the relation with the ratio of star-formation-to-mass surface density, which eliminates this dependence to some degree. We indeed found a moderate positive correlation. Such a relation can be explained if the thick disc is in hydrostatic equilibrium with a pressure contribution from cosmic rays (Sect.\,\ref{ss:hydrostatic_equilibium}).

We then explored the relations between deprojected mean intensities and SFR surface density. We found that only the thin disc shows a strong correlation where the slope is in good agreement with a linear correlation as observed for the global relation at gigahertz frequencies. In contrast, the thick disc shows only a moderate correlation. The fact that the radio continuum emission is only weakly dependent on star formation can be explained by cosmic-ray transport. Since the halo contains mostly old cosmic rays, we do not expect a strong correlation any more with the sites of cosmic ray injection. These results are in good agreement with what is found in studies of edge-on galaxies at kiloparsec resolution where the star-forming areas show a linear relation, whereas the inter-arm areas show a sub-linear relation \citep[e.g.][]{heesen_24b}.  When observing galaxies face-on one needs to take the contribution from the halo into account, particularly when considering that on average the mean deprojected intensity of the thick disc similar to that of the thin disc (Sect.\,\ref{ss:deprojected_intensities}). 

Our results suggest that thick, gaseous discs are in hydrostatic equilibrium supported by cosmic ray pressure \citep{boulares_90a, cox_05a}. Increasing pressure-to-density ratio, such as is potentially expected in areas of higher star-formation rate surface densities, leads to an inflated disc. Contrary, higher mass surface densities lead to more compressed discs. Of course, a hydrostatic equilibrium does not apply globally. We also expect effects of cosmic-ray transport. Finally, we explored normalised scale heights, where we divided them by the effective radius as measured from radio continuum emission. We found a strong positive correlation between the normalised scale heights and the SFR surface density. However, we found no correlation between mass surface density or SFR-to-mass ratio. This suggests that galaxies with high star-formation rate surface densities have a rather roundish shape, whereas galaxies with little star formation only have relatively small vertical extent. Such a correlation can be explained by cosmic-ray transport, as the density scales with the size of the injection region and it relative distance to it (Sect.\,\ref{ss:cosmic_ray_diffusion}).

In the future, it would be worthwhile to extend this kind of study to other frequencies. This may shed some light on the question to what extent the age of the cosmic-ray electrons is different between thin and thick discs. This may then allow us to study the effects of cosmic ray transport in a direct way.

\begin{acknowledgement}
We thank the anonymous referee for a concise and very helpful review. M.S. and R.-J.~D. acknowledge funding from the German Science Foundation DFG, within the Collaborative Research Center SFB1491 “Cosmic Interacting Matters – From Source to Signal. MB acknowledges funding by the Deutsche Forschungsgemeinschaft (DFG) under Germany's Excellence Strategy -- EXC 2121 ``Quantum Universe" --  390833306 and the DFG Research Group "Relativistic Jets". TW acknowledges financial support from the grant CEX2021-001131-S funded by MICIU/AEI/ 10.13039/501100011033, from the coordination of the participation in SKA-SPAIN, funded by the Ministry of Science, Innovation and Universities (MICIU). This research made use of following software packages and other resources: Aladin sky atlas developed at CDS, Strasbourg Observatory, France \citep{bonnarel_00a,boch_14a}; {\sc Astropy} \citep{astropy_13a,astropy_18a};  HyperLeda \citep[\href{http://leda.univ-lyon1.fr}{http://leda.univ-lyon1.fr};][]{makarov_14a}; NASA/IPAC Extragalactic Database (NED), which is operated by the Jet Propulsion Laboratory, California Institute of Technology, under contract with the National Aeronautics and Space Administration; SAOImage DS9 \citep{joye_03a}; and {\sc SciPy} \citep[\href{https://scipy.org}{https://scipy.org};][]{scipy_20a}.
\end{acknowledgement}

\bibliographystyle{aa}
\bibliography{review} 

\begin{thebibliography}{61}
\expandafter\ifx\csname natexlab\endcsname\relax\def\natexlab#1{#1}\fi

\bibitem[{{Ackermann} {et~al.}(2012){Ackermann}, {Ajello}, {Allafort}, {Baldini}, {Ballet}, {Bastieri}, {Bechtol}, {Bellazzini}, {Berenji}, {Bloom}, {Bonamente}, {Borgland}, {Bouvier}, {Bregeon}, {Brigida}, {Bruel}, {Buehler}, {Buson}, {Caliandro}, {Cameron}, {Caraveo}, {Casandjian}, {Cecchi}, {Charles}, {Chekhtman}, {Cheung}, {Chiang}, {Cillis}, {Ciprini}, {Claus}, {Cohen-Tanugi}, {Conrad}, {Cutini}, {de Palma}, {Dermer}, {Digel}, {Silva}, {Drell}, {Drlica-Wagner}, {Favuzzi}, {Fegan}, {Fortin}, {Fukazawa}, {Funk}, {Fusco}, {Gargano}, {Gasparrini}, {Germani}, {Giglietto}, {Giordano}, {Glanzman}, {Godfrey}, {Grenier}, {Guiriec}, {Gustafsson}, {Hadasch}, {Hayashida}, {Hays}, {Hughes}, {J{\'o}hannesson}, {Johnson}, {Kamae}, {Katagiri}, {Kataoka}, {Kn{\"o}dlseder}, {Kuss}, {Lande}, {Longo}, {Loparco}, {Lott}, {Lovellette}, {Lubrano}, {Madejski}, {Martin}, {Mazziotta}, {McEnery}, {Michelson}, {Mizuno}, {Monte}, {Monzani}, {Morselli}, {Moskalenko}, {Murgia}, {Nishino}, {Norris}, {Nuss}, {Ohno}, {Ohsugi}, {Okumura},
  {Omodei}, {Orlando}, {Ozaki}, {Parent}, {Persic}, {Pesce-Rollins}, {Petrosian}, {Pierbattista}, {Piron}, {Pivato}, {Porter}, {Rain{\`o}}, {Rando}, {Razzano}, {Reimer}, {Reimer}, {Ritz}, {Roth}, {Sbarra}, {Sgr{\`o}}, {Siskind}, {Spandre}, {Spinelli}, {Stawarz}, {Strong}, {Takahashi}, {Tanaka}, {Thayer}, {Tibaldo}, {Tinivella}, {Torres}, {Tosti}, {Troja}, {Uchiyama}, {Vandenbroucke}, {Vianello}, {Vitale}, {Waite}, {Wood}, \& {Yang}}]{2012ApJ...755..164A}
{Ackermann}, M., {Ajello}, M., {Allafort}, A., {et~al.} 2012, \apj, 755, 164

\bibitem[{{Astropy Collaboration} {et~al.}(2018){Astropy Collaboration}, {Price-Whelan}, {Sip{\H{o}}cz}, {G{\"u}nther}, {Lim}, {Crawford}, {Conseil}, {Shupe}, {Craig}, {Dencheva}, {Ginsburg}, {VanderPlas}, {Bradley}, {P{\'e}rez-Su{\'a}rez}, {de Val-Borro}, {Aldcroft}, {Cruz}, {Robitaille}, {Tollerud}, {Ardelean}, {Babej}, {Bach}, {Bachetti}, {Bakanov}, {Bamford}, {Barentsen}, {Barmby}, {Baumbach}, {Berry}, {Biscani}, {Boquien}, {Bostroem}, {Bouma}, {Brammer}, {Bray}, {Breytenbach}, {Buddelmeijer}, {Burke}, {Calderone}, {Cano Rodr{\'\i}guez}, {Cara}, {Cardoso}, {Cheedella}, {Copin}, {Corrales}, {Crichton}, {D'Avella}, {Deil}, {Depagne}, {Dietrich}, {Donath}, {Droettboom}, {Earl}, {Erben}, {Fabbro}, {Ferreira}, {Finethy}, {Fox}, {Garrison}, {Gibbons}, {Goldstein}, {Gommers}, {Greco}, {Greenfield}, {Groener}, {Grollier}, {Hagen}, {Hirst}, {Homeier}, {Horton}, {Hosseinzadeh}, {Hu}, {Hunkeler}, {Ivezi{\'c}}, {Jain}, {Jenness}, {Kanarek}, {Kendrew}, {Kern}, {Kerzendorf}, {Khvalko}, {King}, {Kirkby}, {Kulkarni},
  {Kumar}, {Lee}, {Lenz}, {Littlefair}, {Ma}, {Macleod}, {Mastropietro}, {McCully}, {Montagnac}, {Morris}, {Mueller}, {Mumford}, {Muna}, {Murphy}, {Nelson}, {Nguyen}, {Ninan}, {N{\"o}the}, {Ogaz}, {Oh}, {Parejko}, {Parley}, {Pascual}, {Patil}, {Patil}, {Plunkett}, {Prochaska}, {Rastogi}, {Reddy Janga}, {Sabater}, {Sakurikar}, {Seifert}, {Sherbert}, {Sherwood-Taylor}, {Shih}, {Sick}, {Silbiger}, {Singanamalla}, {Singer}, {Sladen}, {Sooley}, {Sornarajah}, {Streicher}, {Teuben}, {Thomas}, {Tremblay}, {Turner}, {Terr{\'o}n}, {van Kerkwijk}, {de la Vega}, {Watkins}, {Weaver}, {Whitmore}, {Woillez}, {Zabalza}, \& {Astropy Contributors}}]{astropy_18a}
{Astropy Collaboration}, {Price-Whelan}, A.~M., {Sip{\H{o}}cz}, B.~M., {et~al.} 2018, \aj, 156, 123

\bibitem[{{Astropy Collaboration} {et~al.}(2013){Astropy Collaboration}, {Robitaille}, {Tollerud}, {Greenfield}, {Droettboom}, {Bray}, {Aldcroft}, {Davis}, {Ginsburg}, {Price-Whelan}, {Kerzendorf}, {Conley}, {Crighton}, {Barbary}, {Muna}, {Ferguson}, {Grollier}, {Parikh}, {Nair}, {Unther}, {Deil}, {Woillez}, {Conseil}, {Kramer}, {Turner}, {Singer}, {Fox}, {Weaver}, {Zabalza}, {Edwards}, {Azalee Bostroem}, {Burke}, {Casey}, {Crawford}, {Dencheva}, {Ely}, {Jenness}, {Labrie}, {Lim}, {Pierfederici}, {Pontzen}, {Ptak}, {Refsdal}, {Servillat}, \& {Streicher}}]{astropy_13a}
{Astropy Collaboration}, {Robitaille}, T.~P., {Tollerud}, E.~J., {et~al.} 2013, \aap, 558, A33

\bibitem[{{Ballet} {et~al.}(2023){Ballet}, {Bruel}, {Burnett}, {Lott}, \& {The Fermi-LAT collaboration}}]{2023arXiv230712546B}
{Ballet}, J., {Bruel}, P., {Burnett}, T.~H., {Lott}, B., \& {The Fermi-LAT collaboration}. 2023, arXiv e-prints, arXiv:2307.12546

\bibitem[{{Boch} \& {Fernique}(2014)}]{boch_14a}
{Boch}, T. \& {Fernique}, P. 2014, in Astronomical Society of the Pacific Conference Series, Vol. 485, Astronomical Data Analysis Software and Systems XXIII, ed. N.~{Manset} \& P.~{Forshay}, 277

\bibitem[{{Bonnarel} {et~al.}(2000){Bonnarel}, {Fernique}, {Bienaym{\'e}}, {Egret}, {Genova}, {Louys}, {Ochsenbein}, {Wenger}, \& {Bartlett}}]{bonnarel_00a}
{Bonnarel}, F., {Fernique}, P., {Bienaym{\'e}}, O., {et~al.} 2000, \aaps, 143, 33

\bibitem[{{Boulares} \& {Cox}(1990)}]{boulares_90a}
{Boulares}, A. \& {Cox}, D.~P. 1990, \apj, 365, 544

\bibitem[{{Cox}(2005)}]{cox_05a}
{Cox}, D.~P. 2005, \araa, 43, 337

\bibitem[{{Dumke} {et~al.}(1995){Dumke}, {Krause}, {Wielebinski}, \& {Klein}}]{dumke_95a}
{Dumke}, M., {Krause}, M., {Wielebinski}, R., \& {Klein}, U. 1995, \aap, 302, 691

\bibitem[{{Ferri{\`e}re}(2001)}]{ferriere_01a}
{Ferri{\`e}re}, K.~M. 2001, Rev.\ Modern Phys., 73, 1031

\bibitem[{{Gajovi{\'c}} {et~al.}(2024){Gajovi{\'c}}, {Adebahr}, {Basu}, {Heesen}, {Br{\"u}ggen}, {de Gasperin}, {Lara-Lopez}, {Oonk}, {Edler}, {Bomans}, {Paladino}, {Gardu{\~n}o}, {L{\'o}pez-Cruz}, {Stein}, {Fritz}, {Piotrowska}, \& {Sinha}}]{gajovic_24a}
{Gajovi{\'c}}, L., {Adebahr}, B., {Basu}, A., {et~al.} 2024, \aap, 689, A68

\bibitem[{{Galante} {et~al.}(2024){Galante}, {Saponara}, {Romero}, \& {Benaglia}}]{galante_24a}
{Galante}, C.~A., {Saponara}, J., {Romero}, G.~E., \& {Benaglia}, P. 2024, \aap, 685, A157

\bibitem[{{Girichidis} {et~al.}(2018){Girichidis}, {Naab}, {Hanasz}, \& {Walch}}]{girichidis_18a}
{Girichidis}, P., {Naab}, T., {Hanasz}, M., \& {Walch}, S. 2018, \mnras, 479, 3042

\bibitem[{{Heald} {et~al.}(2022){Heald}, {Heesen}, {Sridhar}, {Beck}, {Bomans}, {Br{\"u}ggen}, {Chy{\.z}y}, {Damas-Segovia}, {Dettmar}, {English}, {Henriksen}, {Ideguchi}, {Irwin}, {Krause}, {Li}, {Murphy}, {Nikiel-Wroczy{\'n}ski}, {Piotrowska}, {Rand}, {Shimwell}, {Stein}, {Vargas}, {Wang}, {van Weeren}, \& {Wiegert}}]{heald_22a}
{Heald}, G.~H., {Heesen}, V., {Sridhar}, S.~S., {et~al.} 2022, \mnras, 509, 658

\bibitem[{{Heesen} {et~al.}(2019){Heesen}, {Buie}, {Huff}, {Perez}, {Woolsey}, {Rafferty}, {Basu}, {Beck}, {Brinks}, {Horellou}, {Scannapieco}, {Br{\"u}ggen}, {Dettmar}, {Sendlinger}, {Nikiel-Wroczy{\'n}ski}, {Chy{\.z}y}, {Best}, {Heald}, \& {Paladino}}]{heesen_19a}
{Heesen}, V., {Buie}, E., I., {Huff}, C.~J., {et~al.} 2019, \aap, 622, A8

\bibitem[{{Heesen} {et~al.}(2023){Heesen}, {Klocke}, {Br{\"u}ggen}, {Tabatabaei}, {Basu}, {Beck}, {Drabent}, {Nikiel-Wroczy{\'n}ski}, {Paladino}, {Schulz}, \& {Stein}}]{heesen_23a}
{Heesen}, V., {Klocke}, T.~L., {Br{\"u}ggen}, M., {et~al.} 2023, \aap, 669, A8

\bibitem[{{Heesen} {et~al.}(2024{\natexlab{a}}){Heesen}, {Schulz}, {Br{\"u}ggen}, {Edler}, {Stein}, {Paladino}, {Boselli}, {Ignesti}, {Fossati}, \& {Dettmar}}]{heesen_24a}
{Heesen}, V., {Schulz}, S., {Br{\"u}ggen}, M., {et~al.} 2024{\natexlab{a}}, \aap, 682, A83

\bibitem[{{Heesen} {et~al.}(2022){Heesen}, {Staffehl}, {Basu}, {Beck}, {Stein}, {Tabatabaei}, {Hardcastle}, {Chy{\.z}y}, {Shimwell}, {Adebahr}, {Beswick}, {Bomans}, {Botteon}, {Brinks}, {Br{\"u}ggen}, {Dettmar}, {Drabent}, {de Gasperin}, {G{\"u}rkan}, {Heald}, {Horellou}, {Nikiel-Wroczynski}, {Paladino}, {Piotrowska}, {R{\"o}ttgering}, {Smith}, \& {Tasse}}]{heesen_22a}
{Heesen}, V., {Staffehl}, M., {Basu}, A., {et~al.} 2022, \aap, 664, A83

\bibitem[{{Heesen} {et~al.}(2024{\natexlab{b}}){Heesen}, {Wiegert}, {Irwin}, {Crocker}, {Kiehn}, {Li}, {Wang}, {Stein}, {Dettmar}, {Soida}, {Henriksen}, {Gajovi{\'c}}, {Yang}, \& {Br{\"u}ggen}}]{heesen_24b}
{Heesen}, V., {Wiegert}, T., {Irwin}, J., {et~al.} 2024{\natexlab{b}}, \aap, 691, A273

\bibitem[{{Henriksen} \& {Irwin}(2021)}]{henriksen_21}
{Henriksen}, R.~N. \& {Irwin}, J. 2021, \apj, 920, 133

\bibitem[{{Irwin} {et~al.}(2012{\natexlab{a}}){Irwin}, {Beck}, {Benjamin}, {Dettmar}, {English}, {Heald}, {Henriksen}, {Johnson}, {Krause}, {Li}, {Miskolczi}, {Mora}, {Murphy}, {Oosterloo}, {Porter}, {Rand}, {Saikia}, {Schmidt}, {Strong}, {Walterbos}, {Wang}, \& {Wiegert}}]{irwin_12a}
{Irwin}, J., {Beck}, R., {Benjamin}, R.~A., {et~al.} 2012{\natexlab{a}}, \aj, 144, 43

\bibitem[{{Irwin} {et~al.}(2012{\natexlab{b}}){Irwin}, {Beck}, {Benjamin}, {Dettmar}, {English}, {Heald}, {Henriksen}, {Johnson}, {Krause}, {Li}, {Miskolczi}, {Mora}, {Murphy}, {Oosterloo}, {Porter}, {Rand}, {Saikia}, {Schmidt}, {Strong}, {Walterbos}, {Wang}, \& {Wiegert}}]{irwin_12b}
{Irwin}, J., {Beck}, R., {Benjamin}, R.~A., {et~al.} 2012{\natexlab{b}}, \aj, 144, 44

\bibitem[{{Irwin} {et~al.}(2024{\natexlab{a}}){Irwin}, {Beck}, {Cook}, {Dettmar}, {English}, {Heesen}, {Henriksen}, {Jiang}, {Li}, {Lu}, {Mele}, {M{\"u}ller}, {Murphy}, {Porter}, {Rand}, {Skeggs}, {Stein}, {Stein}, {Stil}, {Strong}, {Walterbos}, {Wang}, {Wiegert}, \& {Yang}}]{irwin_24a}
{Irwin}, J., {Beck}, R., {Cook}, T., {et~al.} 2024{\natexlab{a}}, Galaxies, 12, 22

\bibitem[{{Irwin} {et~al.}(2024{\natexlab{b}}){Irwin}, {Cook}, {Stein}, {Dettmar}, {Heesen}, {Wang}, {Wiegert}, {Stein}, \& {Vargas}}]{irwin_24b}
{Irwin}, J., {Cook}, T., {Stein}, M., {et~al.} 2024{\natexlab{b}}, \aj, 168, 138

\bibitem[{{Irwin} {et~al.}(2019){Irwin}, {Wiegert}, {Merritt}, {We{\.z}gowiec}, {Hunt}, {Woodfinden}, {Stein}, {Damas-Segovia}, {Li}, {Wang}, {Johnson}, {Krause}, {Dettmar}, {Im}, {Schmidt}, {Miskolczi}, {Braun}, {Saikia}, {English}, \& {Richardson}}]{irwin_19a}
{Irwin}, J., {Wiegert}, T., {Merritt}, A., {et~al.} 2019, \aj, 158, 21

\bibitem[{{Joye} \& {Mandel}(2003)}]{joye_03a}
{Joye}, W.~A. \& {Mandel}, E. 2003, in Astronomical Society of the Pacific Conference Series, Vol. 295, Astronomical Data Analysis Software and Systems XII, ed. H.~E. {Payne}, R.~I. {Jedrzejewski}, \& R.~N. {Hook}, 489

\bibitem[{{Kennicutt} \& {Evans}(2012)}]{kennicutt_12a}
{Kennicutt}, R.~C. \& {Evans}, N.~J. 2012, \araa, 50, 531

\bibitem[{{Krause} {et~al.}(2020){Krause}, {Irwin}, {Schmidt}, {Stein}, {Miskolczi}, {Carolina Mora-Partiarroyo}, {Wiegert}, {Beck}, {Stil}, {Heald}, {Li}, {Damas-Segovia}, {Vargas}, {Rand}, {West}, {Walterbos}, {Dettmar}, {English}, \& {Woodfinden}}]{krause_20a}
{Krause}, M., {Irwin}, J., {Schmidt}, P., {et~al.} 2020, \aap, 639, A112

\bibitem[{{Krause} {et~al.}(2018){Krause}, {Irwin}, {Wiegert}, {Miskolczi}, {Damas-Segovia}, {Beck}, {Li}, {Heald}, {M{\"u}ller}, {Stein}, {Rand}, {Heesen}, {Walterbos}, {Dettmar}, {Vargas}, {English}, \& {Murphy}}]{krause_18a}
{Krause}, M., {Irwin}, J., {Wiegert}, T., {et~al.} 2018, \aap, 611, A72

\bibitem[{{Li} {et~al.}(2016){Li}, {Beck}, {Dettmar}, {Heald}, {Irwin}, {Johnson}, {Kepley}, {Krause}, {Murphy}, {Orlando}, {Rand}, {Strong}, {Vargas}, {Walterbos}, {Wang}, \& {Wiegert}}]{li_16a}
{Li}, J.-T., {Beck}, R., {Dettmar}, R.-J., {et~al.} 2016, \mnras, 456, 1723

\bibitem[{{Li} {et~al.}(2008){Li}, {Li}, {Wang}, {Irwin}, \& {Rossa}}]{li_08a}
{Li}, J.-T., {Li}, Z., {Wang}, Q.~D., {Irwin}, J.~A., \& {Rossa}, J. 2008, \mnras, 390, 59

\bibitem[{{Li} {et~al.}(2024{\natexlab{a}}){Li}, {Lu}, {Qu}, {Benjamin}, {Bregman}, {Dettmar}, {English}, {Fang}, {Irwin}, {Jiang}, {Li}, {Liu}, {Martini}, {Rand}, {Stein}, {Strong}, {Vargas}, {Wang}, {Wang}, {Wiegert}, {Xu}, \& {Yang}}]{li_24b}
{Li}, J.-T., {Lu}, L.-Y., {Qu}, Z., {et~al.} 2024{\natexlab{a}}, \apj, 967, 78

\bibitem[{{Li} {et~al.}(2024{\natexlab{b}}){Li}, {Sun}, {Ji}, \& {Yang}}]{li_24a}
{Li}, J.-T., {Sun}, W., {Ji}, L., \& {Yang}, Y. 2024{\natexlab{b}}, \apj, 966, 239

\bibitem[{{Lu} {et~al.}(2023){Lu}, {Li}, {Vargas}, {Beck}, {Bregman}, {Dettmar}, {English}, {Fang}, {Heald}, {Li}, {Qu}, {Rand}, {Stein}, {Wang}, {Wang}, {Wiegert}, \& {Zheng}}]{lu_23a}
{Lu}, L.-Y., {Li}, J.-T., {Vargas}, C.~J., {et~al.} 2023, \mnras, 519, 6098

\bibitem[{{Makarov} {et~al.}(2014){Makarov}, {Prugniel}, {Terekhova}, {Courtois}, \& {Vauglin}}]{makarov_14a}
{Makarov}, D., {Prugniel}, P., {Terekhova}, N., {Courtois}, H., \& {Vauglin}, I. 2014, \aap, 570, A13

\bibitem[{{Miskolczi} {et~al.}(2019){Miskolczi}, {Heesen}, {Horellou}, {Bomans}, {Beck}, {Heald}, {Dettmar}, {Blex}, {Nikiel-Wroczy{\'n}ski}, {Chy{\.z}y}, {Stein}, {Irwin}, {Shimwell}, \& {Wang}}]{miskolczi_19a}
{Miskolczi}, A., {Heesen}, V., {Horellou}, C., {et~al.} 2019, \aap, 622, A9

\bibitem[{{Mora-Partiarroyo} {et~al.}(2019){Mora-Partiarroyo}, {Krause}, {Basu}, {Beck}, {Wiegert}, {Irwin}, {Henriksen}, {Stein}, {Vargas}, {Heesen}, {Walterbos}, {Rand}, {Heald}, {Li}, {Kamieneski}, \& {English}}]{mora_19a}
{Mora-Partiarroyo}, S.~C., {Krause}, M., {Basu}, A., {et~al.} 2019, \aap, 632, A10

\bibitem[{{M{\"u}ller} {et~al.}(2017){M{\"u}ller}, {Krause}, {Beck}, \& {Schmidt}}]{mueller_17a}
{M{\"u}ller}, P., {Krause}, M., {Beck}, R., \& {Schmidt}, P. 2017, \aap, 606, A41

\bibitem[{{Offringa} {et~al.}(2014){Offringa}, {McKinley}, {Hurley-Walker}, {Briggs}, {Wayth}, {Kaplan}, {Bell}, {Feng}, {Neben}, {Hughes}, {Rhee}, {Murphy}, {Bhat}, {Bernardi}, {Bowman}, {Cappallo}, {Corey}, {Deshpande}, {Emrich}, {Ewall-Wice}, {Gaensler}, {Goeke}, {Greenhill}, {Hazelton}, {Hindson}, {Johnston-Hollitt}, {Jacobs}, {Kasper}, {Kratzenberg}, {Lenc}, {Lonsdale}, {Lynch}, {McWhirter}, {Mitchell}, {Morales}, {Morgan}, {Kudryavtseva}, {Oberoi}, {Ord}, {Pindor}, {Procopio}, {Prabu}, {Riding}, {Roshi}, {Shankar}, {Srivani}, {Subrahmanyan}, {Tingay}, {Waterson}, {Webster}, {Whitney}, {Williams}, \& {Williams}}]{offringa_14a}
{Offringa}, A.~R., {McKinley}, B., {Hurley-Walker}, N., {et~al.} 2014, \mnras, 444, 606

\bibitem[{{Quataert} {et~al.}(2022){Quataert}, {Thompson}, \& {Jiang}}]{quataert_22a}
{Quataert}, E., {Thompson}, T.~A., \& {Jiang}, Y.-F. 2022, \mnras, 510, 1184

\bibitem[{{Ruszkowski} \& {Pfrommer}(2023)}]{ruszkowski_23a}
{Ruszkowski}, M. \& {Pfrommer}, C. 2023, \aapr, 31, 4

\bibitem[{{Salem} \& {Bryan}(2014)}]{salem_14a}
{Salem}, M. \& {Bryan}, G.~L. 2014, \mnras, 437, 3312

\bibitem[{{Schmidt} {et~al.}(2019){Schmidt}, {Krause}, {Heesen}, {Basu}, {Beck}, {Wiegert}, {Irwin}, {Heald}, {Rand}, {Li}, \& {Murphy}}]{schmidt_19a}
{Schmidt}, P., {Krause}, M., {Heesen}, V., {et~al.} 2019, \aap, 632, A12

\bibitem[{{Smith} {et~al.}(2021){Smith}, {Haskell}, {G{\"u}rkan}, {Best}, {Hardcastle}, {Kondapally}, {Williams}, {Duncan}, {Cochrane}, {McCheyne}, {R{\"o}ttgering}, {Sabater}, {Shimwell}, {Tasse}, {Bonato}, {Bondi}, {Jarvis}, {Leslie}, {Prandoni}, \& {Wang}}]{smith_21a}
{Smith}, D.~J.~B., {Haskell}, P., {G{\"u}rkan}, G., {et~al.} 2021, \aap, 648, A6

\bibitem[{{Socrates} {et~al.}(2008){Socrates}, {Davis}, \& {Ramirez-Ruiz}}]{socrates_08a}
{Socrates}, A., {Davis}, S.~W., \& {Ramirez-Ruiz}, E. 2008, \apj, 687, 202

\bibitem[{{Stein} {et~al.}(2023){Stein}, {Heesen}, {Dettmar}, {Stein}, {Br{\"u}ggen}, {Beck}, {Adebahr}, {Wiegert}, {Vargas}, {Bomans}, {Li}, {English}, {Chy{\.z}y}, {Paladino}, {Tabatabaei}, \& {Strong}}]{stein_23a}
{Stein}, M., {Heesen}, V., {Dettmar}, R.~J., {et~al.} 2023, \aap, 670, A158

\bibitem[{{Stein} {et~al.}(2025){Stein}, {Kleimann}, {Adebahr}, {Dettmar}, {Fichtner}, {English}, {Heesen}, {Kamphuis}, {Irwin}, {Mele}, {Bomans}, {Li}, {Skeggs}, {Wang}, \& {Yang}}]{stein_25a}
{Stein}, M., {Kleimann}, J., {Adebahr}, B., {et~al.} 2025, \aap, 696, A112

\bibitem[{{Stein} {et~al.}(2020){Stein}, {Dettmar}, {Beck}, {Irwin}, {Wiegert}, {Miskolczi}, {Wang}, {English}, {Henriksen}, {Radica}, \& {Li}}]{stein_20a}
{Stein}, Y., {Dettmar}, R.~J., {Beck}, R., {et~al.} 2020, \aap, 639, A111

\bibitem[{{Stein} {et~al.}(2019{\natexlab{a}}){Stein}, {Dettmar}, {Irwin}, {Beck}, {We{\.z}gowiec}, {Miskolczi}, {Krause}, {Heesen}, {Wiegert}, {Heald}, {Walterbos}, {Li}, \& {Soida}}]{stein_19a}
{Stein}, Y., {Dettmar}, R.~J., {Irwin}, J., {et~al.} 2019{\natexlab{a}}, \aap, 623, A33

\bibitem[{{Stein} {et~al.}(2019{\natexlab{b}}){Stein}, {Dettmar}, {We{\.z}gowiec}, {Irwin}, {Beck}, {Wiegert}, {Krause}, {Li}, {Heesen}, {Miskolczi}, {MacDonald}, \& {English}}]{stein_19b}
{Stein}, Y., {Dettmar}, R.~J., {We{\.z}gowiec}, M., {et~al.} 2019{\natexlab{b}}, \aap, 632, A13

\bibitem[{Team {et~al.}(2022)Team, Bean, Bhatnagar, Castro, Meyer, Emonts, Garcia, Garwood, Golap, Villalba, Harris, Hayashi, Hoskins, Hsieh, Jagannathan, Kawasaki, Keimpema, Kettenis, Lopez, Marvil, Masters, McNichols, Mehringer, Miel, Moellenbrock, Montesino, Nakazato, Ott, Petry, Pokorny, Raba, Rau, Schiebel, Schweighart, Sekhar, Shimada, Small, Steeb, Sugimoto, Suoranta, Tsutsumi, van Bemmel, Verkouter, Wells, Xiong, Szomoru, Griffith, Glendenning, \& Kern}]{casa_22a}
Team, T.~C., Bean, B., Bhatnagar, S., {et~al.} 2022, Publications of the Astronomical Society of the Pacific, 134, 114501

\bibitem[{{Thompson} \& {Heckman}(2024)}]{thompson_24a}
{Thompson}, T.~A. \& {Heckman}, T.~M. 2024, \araa, 62, 529

\bibitem[{{Tsukui} {et~al.}(2025){Tsukui}, {Wisnioski}, {Bland-Hawthorn}, \& {Freeman}}]{tsukui_25a}
{Tsukui}, T., {Wisnioski}, E., {Bland-Hawthorn}, J., \& {Freeman}, K. 2025, \mnras~in~press, arXiv:2409.15909

\bibitem[{{Tumlinson} {et~al.}(2017){Tumlinson}, {Peeples}, \& {Werk}}]{tumlinson_17a}
{Tumlinson}, J., {Peeples}, M.~S., \& {Werk}, J.~K. 2017, \araa, 55, 389

\bibitem[{{van Haarlem} {et~al.}(2013){van Haarlem}, {Wise}, {Gunst}, {Heald}, {McKean}, {Hessels}, {de Bruyn}, {Nijboer}, {Swinbank}, {Fallows}, {Brentjens}, {Nelles}, {Beck}, {Falcke}, {Fender}, {H{\"o}randel}, {Koopmans}, {Mann}, {Miley}, {R{\"o}ttgering}, {Stappers}, {Wijers}, {Zaroubi}, {van den Akker}, {Alexov}, {Anderson}, {Anderson}, {van Ardenne}, {Arts}, {Asgekar}, {Avruch}, {Batejat}, {B{\"a}hren}, {Bell}, {Bell}, {van Bemmel}, {Bennema}, {Bentum}, {Bernardi}, {Best}, {B{\^i}rzan}, {Bonafede}, {Boonstra}, {Braun}, {Bregman}, {Breitling}, {van de Brink}, {Broderick}, {Broekema}, {Brouw}, {Br{\"u}ggen}, {Butcher}, {van Cappellen}, {Ciardi}, {Coenen}, {Conway}, {Coolen}, {Corstanje}, {Damstra}, {Davies}, {Deller}, {Dettmar}, {van Diepen}, {Dijkstra}, {Donker}, {Doorduin}, {Dromer}, {Drost}, {van Duin}, {Eisl{\"o}ffel}, {van Enst}, {Ferrari}, {Frieswijk}, {Gankema}, {Garrett}, {de Gasperin}, {Gerbers}, {de Geus}, {Grie{\ss}meier}, {Grit}, {Gruppen}, {Hamaker}, {Hassall}, {Hoeft}, {Holties},
  {Horneffer}, {van der Horst}, {van Houwelingen}, {Huijgen}, {Iacobelli}, {Intema}, {Jackson}, {Jelic}, {de Jong}, {Juette}, {Kant}, {Karastergiou}, {Koers}, {Kollen}, {Kondratiev}, {Kooistra}, {Koopman}, {Koster}, {Kuniyoshi}, {Kramer}, {Kuper}, {Lambropoulos}, {Law}, {van Leeuwen}, {Lemaitre}, {Loose}, {Maat}, {Macario}, {Markoff}, {Masters}, {McFadden}, {McKay-Bukowski}, {Meijering}, {Meulman}, {Mevius}, {Middelberg}, {Millenaar}, {Miller-Jones}, {Mohan}, {Mol}, {Morawietz}, {Morganti}, {Mulcahy}, {Mulder}, {Munk}, {Nieuwenhuis}, {van Nieuwpoort}, {Noordam}, {Norden}, {Noutsos}, {Offringa}, {Olofsson}, {Omar}, {Orr{\'u}}, {Overeem}, {Paas}, {Pandey-Pommier}, {Pandey}, {Pizzo}, {Polatidis}, {Rafferty}, {Rawlings}, {Reich}, {de Reijer}, {Reitsma}, {Renting}, {Riemers}, {Rol}, {Romein}, {Roosjen}, {Ruiter}, {Scaife}, {van der Schaaf}, {Scheers}, {Schellart}, {Schoenmakers}, {Schoonderbeek}, {Serylak}, {Shulevski}, {Sluman}, {Smirnov}, {Sobey}, {Spreeuw}, {Steinmetz}, {Sterks}, {Stiepel}, {Stuurwold},
  {Tagger}, {Tang}, {Tasse}, {Thomas}, {Thoudam}, {Toribio}, {van der Tol}, {Usov}, {van Veelen}, {van der Veen}, {ter Veen}, {Verbiest}, {Vermeulen}, {Vermaas}, {Vocks}, {Vogt}, {de Vos}, {van der Wal}, {van Weeren}, {Weggemans}, {Weltevrede}, {White}, {Wijnholds}, {Wilhelmsson}, {Wucknitz}, {Yatawatta}, {Zarka}, {Zensus}, \& {van Zwieten}}]{vanHaarlem_13a}
{van Haarlem}, M.~P., {Wise}, M.~W., {Gunst}, A.~W., {et~al.} 2013, \aap, 556, A2

\bibitem[{{Vargas} {et~al.}(2019){Vargas}, {Walterbos}, {Rand}, {Stil}, {Krause}, {Li}, {Irwin}, \& {Dettmar}}]{vargas_19a}
{Vargas}, C.~J., {Walterbos}, R. A.~M., {Rand}, R.~J., {et~al.} 2019, \apj, 881, 26

\bibitem[{Virtanen {et~al.}(2020)Virtanen, Gommers, Oliphant, Haberland, Reddy, Cournapeau, Burovski, Peterson, Weckesser, Bright, {van der Walt}, Brett, Wilson, Millman, Mayorov, Nelson, Jones, Kern, Larson, Carey, Polat, Feng, Moore, {VanderPlas}, Laxalde, Perktold, Cimrman, Henriksen, Quintero, Harris, Archibald, Ribeiro, Pedregosa, {van Mulbregt}, \& {SciPy 1.0 Contributors}}]{scipy_20a}
Virtanen, P., Gommers, R., Oliphant, T.~E., {et~al.} 2020, Nature Methods, 17, 261

\bibitem[{{Wang} {et~al.}(2001){Wang}, {Immler}, {Walterbos}, {Lauroesch}, \& {Breitschwerdt}}]{wang_01a}
{Wang}, Q.~D., {Immler}, S., {Walterbos}, R., {Lauroesch}, J.~T., \& {Breitschwerdt}, D. 2001, \apjl, 555, L99

\bibitem[{{Werhahn} {et~al.}(2021){Werhahn}, {Pfrommer}, {Girichidis}, {Puchwein}, \& {Pakmor}}]{werhahn_21a}
{Werhahn}, M., {Pfrommer}, C., {Girichidis}, P., {Puchwein}, E., \& {Pakmor}, R. 2021, \mnras, 505, 3273

\bibitem[{{Wiegert} {et~al.}(2015){Wiegert}, {Irwin}, {Miskolczi}, {Schmidt}, {Mora}, {Damas-Segovia}, {Stein}, {English}, {Rand}, {Santistevan}, {Walterbos}, {Krause}, {Beck}, {Dettmar}, {Kepley}, {Wezgowiec}, {Wang}, {Heald}, {Li}, {MacGregor}, {Johnson}, {Strong}, {DeSouza}, \& {Porter}}]{wiegert_15a}
{Wiegert}, T., {Irwin}, J., {Miskolczi}, A., {et~al.} 2015, \aj, 150, 81

\bibitem[{{Xu} {et~al.}(2025){Xu}, {Yang}, {Li}, {Liu}, {Irwin}, {Dettmar}, {Stein}, {Wiegert}, {Wang}, \& {English}}]{xu_25a}
{Xu}, J., {Yang}, Y., {Li}, J.-T., {et~al.} 2025, \apj, 978, 5

\end{thebibliography}

\appendix

\onecolumn

\section{Galaxy sample information}
\label{as:sample}

In Table\,\ref{tab:sample} we present some general information about our sample galaxies. 

\begin{table*}[h!]
\centering
\caption{Properties of galaxies in the sample.}
\label{tab:sample}
\begin{tabular}{l ccccc ccc}
\hline\hline
Galaxy & $d$ & $\log_{10} r_\star$ & $\log_{10}$ \sfrd & $\log_{10}$ \msd  & $\log_{10}$ (\sfrd/\msd) & SFR & $i$ & pa  \\
& (Mpc) & (kpc) & (\usfrd) & (\umsd) & (\ussfr) & (\usfr) & $(\degr)$ & $(\degr)$ \\  
(1) & (2)  & (3) & (4) & (5) & (6) &  (7) & (8) & (9)\\ \hline
N0891 & $9.1$ & $1.10$ & $-2.42$ & $8.39$ & $-10.81$ & $1.88$ & $84$ & $22$ \\ 
N2683 & $6.27$ & $0.67$ & $-2.44$ & $8.72$ & $-11.16$ & $0.25$ & $79$ & $44$ \\ 
N2820 & $26.5$ & $0.84$ & $-2.04$ & $8.34$ & $-10.38$ & $1.35$ & $88$ & $65$ \\ 
N3003 & $25.4$ & $1.14$ & $-2.59$ & $7.77$ & $-10.35$ & $1.56$ & $85$ & $79$ \\ 
N3044 & $20.3$ & $0.96$ & $-2.17$ & $8.24$ & $-10.41$ & $1.75$ & $85$ & $-67$ \\ 
N3079 & $20.6$ & $1.11$ & $-2.02$ & $8.30$ & $-10.31$ & $5.08$ & $88$ & $-14$ \\ 
N3432 & $9.42$ & $0.70$ & $-2.18$ & $8.17$ & $-10.35$ & $0.51$ & $85$ & $41$ \\ 
N3448 & $24.5$ & $0.79$ & $-1.84$ & $8.42$ & $-10.25$ & $1.78$ & $78$ & $65$ \\ 
N3556 & $14.09$ & $1.10$ & $-2.14$ & $7.98$ & $-10.12$ & $3.57$ & $81$ & $79$ \\ 
N3628 & $8.5$ & $1.09$ & $-2.52$ & $8.32$ & $-10.85$ & $1.41$ & $87$ & $-76$ \\ 
N3735 & $42.0$ & $1.24$ & $-2.18$ & $8.26$ & $-10.43$ & $6.23$ & $85$ & $-50$ \\ 
N4013 & $16.0$ & $0.90$ & $-2.46$ & $8.35$ & $-10.80$ & $0.71$ & $88$ & $65$ \\ 
N4096 & $10.32$ & $0.77$ & $-2.19$ & $8.34$ & $-10.53$ & $0.71$ & $82$ & $20$ \\ 
N4157 & $15.6$ & $0.92$ & $-2.09$ & $8.38$ & $-10.47$ & $1.76$ & $83$ & $65$ \\ 
N4217 & $20.6$ & $1.07$ & $-2.36$ & $8.28$ & $-10.64$ & $1.89$ & $86$ & $50$ \\ 
N4302 & $19.41$ & $1.03$ & $-2.60$ & $8.25$ & $-10.85$ & $0.92$ & $90$ & $-2$ \\ 
N4388 & $16.6$ & $0.74$ & $-1.60$ & $8.57$ & $-10.17$ & $2.42$ & $79$ & $-89$ \\ 
N4565 & $11.9$ & $1.26$ & $-3.03$ & $7.89$ & $-10.92$ & $0.96$ & $86$ & $-45$ \\ 
N4631 & $7.4$ & $1.07$ & $-2.21$ & $7.98$ & $-10.19$ & $2.62$ & $85$ & $86$ \\ 
N4666 & $27.5$ & $1.21$ & $-1.89$ & $8.20$ & $-10.09$ & $10.5$ & $76$ & $41$ \\ 
N5775 & $28.9$ & $1.20$ & $-2.03$ & $8.17$ & $-10.20$ & $7.56$ & $86$ & $-35$ \\ 
N5907 & $16.8$ & $1.25$ & $-2.66$ & $8.13$ & $-10.79$ & $2.21$ & $90$ & $-24$ \\ 
\hline
\end{tabular}
\tablefoot{Column (2) $d$ is the distance \citep{wiegert_15a}; (3) $r_\star$ is the star formation radius as from near infrared  \citep{wiegert_15a}; (4) is the SFR surface density within $r_\star$; (5) total mass surface density from \citet{irwin_12a} scaled to our distances (6) ratio of SFR-to-mass surface density; (7) star formation rate from H$\alpha$ and mid-infrared  \citep{vargas_19a}; (8) inclination and position angles from \citet[][N2820, 3003, 3044, 3079, 3432, 3735, 3877, 4013, 4157, 4217, 4302, 5775]{krause_18a}, otherwise \citet{irwin_12a}; (9) position angle from \citet[][see inclination angle]{krause_18a}, otherwise Hyperleda}
\end{table*}

\newpage

\section{Observational results}
\label{as:results}

In Table\,\ref{tab:results} we present observational and derived results for our sample galaxies.

\begin{table*}[h!]
\centering
\caption{Scale heights and amplitudes of thin and thick radio discs; effective radii in the radio continuum.}
\label{tab:results}
\begin{tabular}{l cc cc cc c}
\hline\hline
Galaxy & $w_1$ & $w_2$ & $h_1$ & $h_2$ & $r_{\rm e}$ & $h_1$ & $h_2$ \\
& \multicolumn{2}{c}{(\uint)} & (arcsec) & (arcsec) & (arcsec) & (kpc) & (kpc)  \\  \hline
N0891                  & $2.81\pm0.94$ & $0.88\pm0.22$ & $2.86\pm0.26$ & $22.77\pm2.05$ & $132.40\pm4.33 $ & $0.13\pm0.01$ & $1.00\pm0.09$\\ 
N2683                  & $0.26\pm0.04$ & $0.17\pm0.05$ & $8.86\pm4.76$ & $21.64\pm2.52$ & $86.41\pm1.88 $ & $0.27\pm0.14$ & $0.66\pm0.08$\\ 
N2820\tablefootmark{a} & $1.53\pm0.43$ & $0.60\pm0.22$ & $2.86\pm0.66$ & $10.29\pm0.85$ & $30.31\pm1.13 $ & $0.37\pm0.08$ & $1.32\pm0.11$\\ 
N3003                  & $0.85\pm0.05$ & $0.27\pm0.06$ & $2.54\pm0.50$ & $12.20\pm1.52$ & $56.35\pm3.89 $ & $0.31\pm0.06$ & $1.50\pm0.19$\\ 
N3044\tablefootmark{a} & $3.19\pm1.34$ & $0.65\pm0.42$ & $3.69\pm0.95$ & $10.77\pm2.05$ & $24.24\pm1.05 $ & $0.36\pm0.09$ & $1.06\pm0.20$\\ 
N3079\tablefootmark{c} & $4.86\pm0.20$ & $0.28\pm0.22$ & $5.41\pm0.86$ & $13.57\pm4.03$ & $49.58\pm1.87 $ & $0.54\pm0.09$ & $1.36\pm0.40$\\ 
N3432                  & $0.49\pm0.14$ & $0.18\pm0.10$ & $5.83\pm1.88$ & $20.36\pm1.58$ & $76.38\pm3.40 $ & $0.27\pm0.09$ & $0.93\pm0.07$\\ 
N3448                  & $0.88\pm0.18$ & $0.73\pm0.33$ & $3.41\pm1.40$ & $7.96\pm0.84$ & $32.07\pm1.60 $ & $0.41\pm0.17$ & $0.95\pm0.10$\\ 
N3556\tablefootmark{a} & $0.48\pm0.05$ & $0.15\pm0.14$ & $13.05\pm5.19$ & $31.57\pm6.86$ & $125.70\pm11.70 $ & $0.89\pm0.35$ & $2.16\pm0.47$\\ 
N3628\tablefootmark{a} & $2.12\pm0.12$ & $0.24\pm0.14$ & $6.53\pm1.82$ & $23.10\pm5.77$ & $54.55\pm1.79 $ & $0.27\pm0.08$ & $0.95\pm0.24$\\ 
N3735\tablefootmark{c} & $2.37\pm0.88$ & $0.02\pm0.03$ & $5.13\pm0.72$ & $23.62\pm3.54$ & $29.36\pm0.86 $ & $1.04\pm0.15$ & $4.81\pm0.72$\\ 
N4013                  & $0.80\pm0.60$ & $0.13\pm0.02$ & $1.11\pm0.19$ & $10.67\pm0.32$ & $85.87\pm4.46 $ & $0.09\pm0.01$ & $0.83\pm0.02$\\ 
N4096\tablefootmark{b} & $0.57\pm0.09$ & $0.16\pm0.01$ & $2.53\pm0.64$ & $22.24\pm1.53$ & $64.47\pm2.92 $ & $0.13\pm0.03$ & $1.11\pm0.08$\\ 
N4157                  & $0.98\pm0.09$ & $0.10\pm0.10$ & $8.90\pm1.58$ & $17.25\pm4.28$ & $75.38\pm2.54 $ & $0.67\pm0.12$ & $1.30\pm0.32$\\ 
N4217\tablefootmark{a} & $1.18\pm0.17$ & $0.48\pm0.20$ & $2.95\pm0.97$ & $11.29\pm1.95$ & $59.57\pm1.90 $ & $0.29\pm0.10$ & $1.13\pm0.19$\\ 
N4302                  & $0.44\pm0.15$ & $0.17\pm0.04$ & $1.59\pm0.30$ & $9.90\pm0.89$ & $73.22\pm1.81 $ & $0.15\pm0.03$ & $0.93\pm0.08$\\ 
N4388                  & $2.71\pm0.22$ & $0.09\pm0.18$ & $5.07\pm1.17$ & $10.25\pm1.56$ & $42.53\pm2.00 $ & $0.41\pm0.09$ & $0.82\pm0.13$\\ 
N4565                  & $0.22\pm0.01$ & $0.08\pm0.09$ & $10.48\pm3.43$ & $14.38\pm1.02$ & $213.30\pm4.36 $ & $0.60\pm0.20$ & $0.83\pm0.06$\\ 
N4631\tablefootmark{a} & $4.52\pm0.10$ & $1.17\pm0.11$ & $4.90\pm0.56$ & $28.58\pm0.99$ & $100.00\pm10.00 $ & $0.18\pm0.02$ & $1.03\pm0.04$\\ 
N4666                  & $3.62\pm0.28$ & $0.88\pm0.36$ & $6.88\pm1.34$ & $19.14\pm2.61$ & $43.33\pm1.37 $ & $0.92\pm0.18$ & $2.55\pm0.35$\\ 
N5775                  & $1.76\pm0.56$ & $0.29\pm0.06$ & $5.65\pm0.29$ & $22.59\pm4.27$ & $80.08\pm5.18 $ & $0.79\pm0.04$ & $3.17\pm0.60$\\ 
N5907\tablefootmark{a} & $0.41\pm0.08$ & $0.01\pm0.02$ & $10.29\pm1.62$ & $24.78\pm3.62$ & $140.40\pm4.43 $ & $0.84\pm0.13$ & $2.02\pm0.29$\\ 
\hline
\end{tabular}
\tablefoot{Results refer to 3\,GHz radio continuum emission at 7\,arcsec FWHM angular resolution. For NGC\,4631, the effective radius was estimated without fit. \tablefoottext{a}{Both haloes individually fitted;} \tablefoottext{b}{only northern halo ($z>0$);} \tablefoottext{c}{only southern halo ($z<0$).}}
\end{table*}

\end{document}